\documentclass[12pt, draftclsnofoot, onecolumn]{IEEEtran}
\usepackage{graphicx, times, amsmath, amsfonts, amsthm, comment}
\usepackage{amssymb,epstopdf}
\usepackage{algorithmic}
\usepackage{algorithm}
\usepackage{color, soul}
\usepackage{mathtools}
\usepackage{caption}
\usepackage{subcaption}

\newcommand{\beq}{\begin{equation}}
\newcommand{\eeq}{\end{equation}}
\newcommand{\tbf}{\textbf}
\newcommand{\tit}{\textit}

\newcommand{\ud}{\mathrm{d}}

\newcommand*{\mathcolor}{}
\def\mathcolor#1#{\mathcoloraux{#1}}
\newcommand*{\mathcoloraux}[3]{%
  \protect\leavevmode
  \begingroup
    \color#1{#2}#3%
  \endgroup
}

\theoremstyle{plain}
\newtheorem{propcounter}{Proposition}
\newtheorem{proposition}[propcounter]{Proposition}
\theoremstyle{plain}
\newtheorem{corocounter}{Corollary}
\newtheorem{corollary}[corocounter]{Corollary}
\theoremstyle{plain}

\theoremstyle{plain}
\newtheorem{assumecounter}{Assumption}
\newtheorem{assumption}[assumecounter]{Assumption}

\newcommand {\Fcal}{\mathcal{F}}

\newcommand {\Kcal}{\mathcal{K}}

\newcommand {\Ncal}{\mathcal{N}}
\newcommand {\Ocal}{\mathcal{O}}

\begin{document}

\title{Infrastructure Sharing for Mobile Network Operators: Analysis of Trade-offs and Market}

\author{
\IEEEauthorblockN{Tachporn Sanguanpuak\IEEEauthorrefmark{1}, Sudarshan Guruacharya\IEEEauthorrefmark{2}, Ekram Hossain\IEEEauthorrefmark{2},
\\ Nandana Rajatheva\IEEEauthorrefmark{1}, and Matti Latva-aho\IEEEauthorrefmark{1}}
%

\thanks{T. Sanguanpuak, N. Rajatheva, and M. Latva-aho are with the Centre for Wireless Communications (CWC), Dept. of Commun. Eng., University of Oulu, Finland (E-mails: \{tachporn.sanguanpuak, nandana.rajathava, matti.latva-aho\}@oulu.fi); and S. Guruacharya and E. Hossain are with the Department of Electrical and Computer Engineering, University of Manitoba, Canada (E-mail: \{Sudarshan.Guruacharya, Ekram.Hossain\}@umanitoba.ca).}
}
\maketitle

\begin{abstract}

The conflicting problems of growing mobile service demand and underutilization of dedicated spectrum has given rise to a paradigm where mobile network operators (MNOs) share their infrastructure among themselves in order to lower their operational costs, while at the same time increase the usage of their existing network resources. We model and analyze such an infrastructure sharing system considering a single buyer MNO and multiple seller MNOs.  
Assuming that the locations of the BSs can be modeled as a homogeneous Poisson point process, we find the downlink signal-to-interference-plus-noise ratio (SINR) coverage probability for a user served by the buyer MNO in an infrastructure sharing environment. We analyze the trade-off between increasing the transmit power of a BS and the intensity of BSs owned by the buyer MNO required to achieve a given quality-of-service (QoS) in terms of the SINR coverage probability. %
Also, for a seller MNO, we analyze the power consumption of the network per unit area (i.e., areal power consumption) which is shown to be a piecewise continuous function of BS intensity, composed of a linear and a convex function.  Accordingly, the BS intensity of the seller MNO can be optimized to minimize the areal power consumption while achieving a minimum QoS for the buyer MNO. We then use these results to formulate a single-buyer multiple-seller BS infrastructure market. The buyer MNO  is concerned with finding which seller MNO to purchase from and what fraction of BSs to purchase. On the sellers' side, the problem of pricing and determining the fraction of infrastructure to be sold is formulated as a Cournot oligopoly market.  We prove that the iterative update of each seller's best response always converges to the Nash Equilibrium.
\end{abstract}


\section{Introduction} \label{section:introduction}

\subsection{Motivation}
In recent years, the concept of network infrastructure sharing has been investigated to address two kinds of concerns. On one hand, with the growing demand for mobile services, the underutilization of dedicated spectrum auctioned off to the mobile network operators (MNOs) has become a bottleneck for the future growth of the industry \cite{Cisco2014}. While on the other hand, in areas or time periods where demand can be low -- such as in rural areas or developing countries, or during night time -- the high cost of network infrastructure forces the operators to charge higher prices from their customers, making the mobile services unaffordable to many people, hence further driving down the demand \cite{ITU2016,GSMA2012}. As capacity demand is expected to increase, MNOs are required to increase their capital (CAPEX) and operational expenses (OPEX) accordingly. One possible paradigm to address these issues is to allow the MNOs to share their infrastructures in order to maximize the use of existing network resources while simultaneously minimizing the operational costs \cite{Cisco2014, ITU2016, GSMA2012}.

In \cite{3GPP2014}, the third generation partnership project (3GPP) has defined standards for network sharing. Accordingly, by sharing the infrastructure among the MNOs, it also allows for a faster deployment of network services. Such sharing of infrastructure can be \tit{passive} or \tit{active}. \tit{Passive sharing} refers to the sharing of physical space, such as buildings, sites, masts, and power supply. In \tit{active sharing}, active elements of the network such as antennas, backhaul, base stations, and elements of core network are shared. Thus, such active sharing allows mobile roaming, which allows an MNO to make use of another network in a place where it has no coverage or infrastructure of its own. According to a market survey in \cite{MNSR2010}, infrastructure sharing has been deployed by over $65 \%$ of European MNOs, involving both active and passive radio access network (RAN) sharing. This trend is expected to grow in the future. In recent years, the concept of drone base stations (BSs) has been introduced \cite{DroneMag2016}. The concept of sharing can also be extended to drone-based infrastructures. 

\subsection{Related Work and Contribution}
In \cite{5GAmerica}, the neutral host network deployment was proposed where the MNOs deploy cells in the best positions with optimal tuning to satisfy the quality-of-experience (QoE). 
The technical and the financial impact of infrastructure sharing was investigated in \cite{Frisanco2008}. In \cite{Xavier2014}, RAN sharing was considered based on BS virtualization, which allows multiple entities to share the same spectrum. In \cite{Jorswieck2014},  the benefit of inter-operator spectrum sharing was demonstrated. Resource sharing in the context of heterogeneous network and cloud RAN concepts was proposed in \cite{Marcelo2015}. In \cite{Jun2015}, a service-oriented framework for RAN sharing which decouples MNOs from radio resource by providing application-level differentiated services was studied. In \cite{Antonopoulos2015}, the authors studied infrastructure sharing along with BS switch off mechanism. In \cite{Bousia2016}, game theory was applied to study multi-operator infrastructure sharing for BS de-activation. The work in \cite{Frisanco2008}--\cite{Bousia2016} considered deterministic network models.

In \cite{Kibilda2015}, stochastic geometry was used to investigate infrastructure sharing, spectrum sharing, and the combination of two in large-scale cellular networks. When both types of sharing are allowed, the authors showed that a trade-off exists between coverage and data rate performance. In \cite{Kibilda2016}, the point processes that model the spatial characteristics of the BSs belonging to multiple MNOs were empirically studied, using the data from field surveys. 
In \cite{Wang2016}, the authors also exploited stochastic geometry to study the trade-off involved in spectrum sharing and infrastructure sharing.

In this paper, we model and analyze the infrastructure sharing problem in a large-scale cellular network by exploiting tools and results from  stochastic geometry. The tradeoff between transmit power and the intensity of the BS deployment for a buyer MNO and that between the areal power consumption and the BS intensity for the seller MNO are analyzed in an infrastructure sharing scenario. Also, the market competition among MNOs for selling and buying infrastructure is modeled and analyzed. Although the problem of spectrum and infrastructure sharing was considered in \cite{Kibilda2015}, the tradeoffs as well as the market were not analyzed.  Note that in \cite{Sanguanpuak_EuCNC2017}, we modeled and analyzed the problem of spectrum sharing among network operators using a stochastic geometry approach. 

We consider multiple co-located deployment of network infrastructures by different MNOs, where the MNOs are assumed to operate over orthogonal frequency bands. In the infrastructure sharing deployment, each BS can be utilized by the users subscribed to more than one MNO. The MNOs that install the BS are considered as potential sellers of the BS infrastructure (i.e. incumbent MNOs). The entrant MNO that use the BS of the incumbent MNOs to serve their users is considered as the buyer. In the presence of multiple seller MNOs, it is assumed that they compete with each other to sell their infrastructure to a  potential buyer.  Note that our study in this paper focuses only on  infrastructure sharing among the MNOs. We consider that BSs are randomly scattered in two dimensional 2D plane. 
First, we study the strategy of a buyer MNO, that decides which MNOs to buy the infrastructure from, and how much infrastructure to buy from them. We propose a cost minimization problem for the buyer MNO, while guaranteeing the quality-of-service (QoS) to its users, in terms of the SINR coverage probability, as an optimization problem. Next, we propose the market from the point of view of the sellers, which compete with each other to sell the infrastructure. We model the competition among the seller MNOs as a Cournot-Nash game. The seller MNOs compete with each other in terms of their supply (a fraction of infrastructure to be shared), the associated cost (e.g. due to power consumption at the BSs), and the selling price, with the objective of gaining the highest profit. As such we find the Cournot-Nash equilibrium and obtain the equilibrium price. We use results from stochastic geometric analysis of large-scale networks to evaluate SINR outage probability and power consumption to model such a market.


The major contributions of the paper can be summarized as follows:

\begin{itemize}
\item The paper presents an infrastructure sharing model with multiple seller MNOs and single MNO. The downlink SINR coverage probability, which is considered to be the QoS metric for the buyer MNO, is analyzed using stochastic geometry.
\item Subsequently, the trade-off between increasing the transmit power of a BS versus increasing the BS deployment density for the buyer MNO  is analyzed. It is shown that there is an upper bound beyond which increasing the transmit power cannot improve the coverage probability. Infrastructure sharing is beneficial when the QoS is above this bound. Infrastructure sharing can improve the  cellular coverage as long as the BS interference and BS association are decoupled.
\item For a seller MNO, since its profit depends on its cost of network operation, the areal power consumption (i.e., power consumption per unit area) at the BSs is analyzed.
\item The optimal strategy for the buyer MNO, in order to minimize the cost of purchase, is obtained by using Lagrange multiplier method. We use greedy algorithm to find which seller MNO and how much infrastructure to purchase from.
\item The optimal strategy for the seller MNOs, in terms of the fraction of infrastructure to be shared and the pricing for the infrastructure, is obtained by computing the equilibrium of a Cournot-Nash market/game.
\end{itemize}




\subsection{Organization}
The rest of the paper is organized as follows: Section \ref{section:systemmodel} describes the system model and the assumptions. Section \ref{sec:stogeoana} gives the  stochastic geometrical analysis of the downlink SINR coverage probability of a typical user based on two scenarios: (i) all the BSs of the seller MNOs serve UEs subscribing to the buyer MNO (i.e., interference is caused at the reference user from downlink transmissions of \textbf{all} BSs of all of the seller MNOs as well  those from the buyer MNO's) and (ii)  some of the BSs of the seller MNOs serve UEs subscribing to the buyer MNO (i.e., interference is caused at the reference UE from a \textbf{fraction} of all the BSs of the MNOs including  those from the buyer MNO's). The trade-off between transmit power and infrastructure (i.e., intensity of BSs) is analyzed in Section \ref{subsec:Tx_infra}. Section \ref{section:strategybuyer} models the strategic behavior of a buyer MNO when buying infrastructure from multiple seller MNOs. Section \ref{section:cournotseller} analyzes the competition among multiple sellers using a Cournot-Nash game. The numerical results are presented in Section \ref{section:Numerical Results} before the paper is concluded in Section \ref{section:Conclusion}.

\section{System Model and Assumptions} \label{section:systemmodel}
Consider a system with $K+1$ MNOs given by the set $\Kcal = \{0, 1, \ldots, K\}$  that serves a common geographical area. We consider multiple sellers single buyer market for infrastructure sharing. We assume that an MNO cannot be both buyer and seller at the same time. Let MNO-$0$ denote our buyer MNO. Let the set of BSs owned by MNO-$k$ be given by $\Fcal_k$, where $k\in\Kcal$. Each of the BSs and UEs are assumed to be equipped with a single antenna. The maximum transmit power of each BS is $p_{\max}$. Also, a UE subscribed to an MNO associates to the nearest BS belonging to that MNO. The BSs owned by different MNOs are spatially distributed according to homogeneous Poisson point processes (PPPs). Let the spatial intensity of BSs per unit area of MNO-$k$ be denoted by $\lambda_k$, where $k\in\Kcal$. Furthermore, each MNO-$k$, is assumed to operate on orthogonal spectrum. Thus, there is no inter-operator interference among the MNOs. However, since the all the BSs belonging to an MNO utilize a common spectrum, intra-operator interference is present. During the sharing of infrastructure,   the following assumptions hold:

\begin{assumption}
When the buyer MNO-$0$ is allowed to use the infrastructure of a seller MNO-$k$, where $k\in \Kcal\backslash\{0\}$, the typical UE of MNO-$0$ associates with the nearest available BSs owned by MNO-$0$ or the seller MNO-$k$.
\end{assumption}

If the buyer MNO-$0$ shares infrastructure with $\Ncal \subseteq \Kcal\backslash\{0\}$ seller MNOs, then a UE subscribed to MNO-$0$ can effectively associate to any one of the enlarged set of BSs given by $\Fcal = \Fcal_0 \cup (\cup_{k \in \Ncal} \Fcal_k)$. This implies that the net intensity of the BSs that a typical UE of MNO-$0$ can associate itself with is
\beq
\lambda_A = \lambda_0 + \sum_{k \in \Ncal} \lambda_k,
\label{eqn:lambda}
\eeq
due to the superposition property of PPP. In places where it is not ambiguous, we can denote the overall net  intensity of the BSs of all MNOs as sum of all $\lambda_k$ by,
\[\lambda = \lambda_0 + \sum_{k \in \Ncal} \lambda_k. \]
\begin{assumption}
The buyer MNO-$0$ is assumed to use the infrastructure, but \tbf{not} the spectrum, belonging to a seller MNO-$k$, where $k\in \Kcal\backslash\{0\}$. As such a UE of MNO-$0$ served by the shared BS of a seller MNO-$k$ has to operate on the spectrum belonging to the MNO-$0$ itself. We will consider two possible cases for the interference experienced by the typical UE of MNO-$0$:
\begin{enumerate}
\item When every shared BS of the seller MNOs serves a user from MNO-$0$, we have the intensity of interfering BSs as
\begin{equation}
\lambda_I = \lambda_0 +  \sum_{k \in \Ncal}\lambda_k = \lambda_A
\label{eqn:lambdaI_assump2.1}
\end{equation}
\item When only some of the BSs of the seller MNOs serve users from MNO-$0$, the intensity of interfering BSs is given by
\begin{equation}
\lambda_I = \sum_{k \in \Ncal \cup \{0\}} w_k \lambda_k
\label{eqn:lambdaI_assump2.2}
\end{equation}
where $w_k$ denotes the level of activity of UE of MNO-$0$ using infrastructure of seller MNO-$k$, such that $k \in \Ncal$,  and $\sum_{k \in \Ncal \cup \{0\}} w_k = 1$. In this case, $\lambda_A$ is given by (\ref{eqn:lambda}).
\end{enumerate}
\end{assumption}

Note that despite the sharing of BSs among MNOs, there is no inter-operator interference among MNOs in our system model, since each MNO operates over a different spectrum. Due to \tbf{Assumption 2}, the buyer will purchase only the infrastructure of the seller MNOs and not the spectrum. \tbf{Assumption 2.1} is a worst case assumption, while \tbf{Assumption 2.2} is a more realistic assumption. While in \tbf{Assumption 2.1} $\lambda_I = \lambda_A$, in \tbf{Assumption 2.2} $\lambda_I$ and $\lambda_A$ are {\em de-coupled}.

\begin{figure}[h]
    \centering
    \begin{subfigure}[h]{0.5\textwidth}
        \centering
        \includegraphics[height=2.1in]{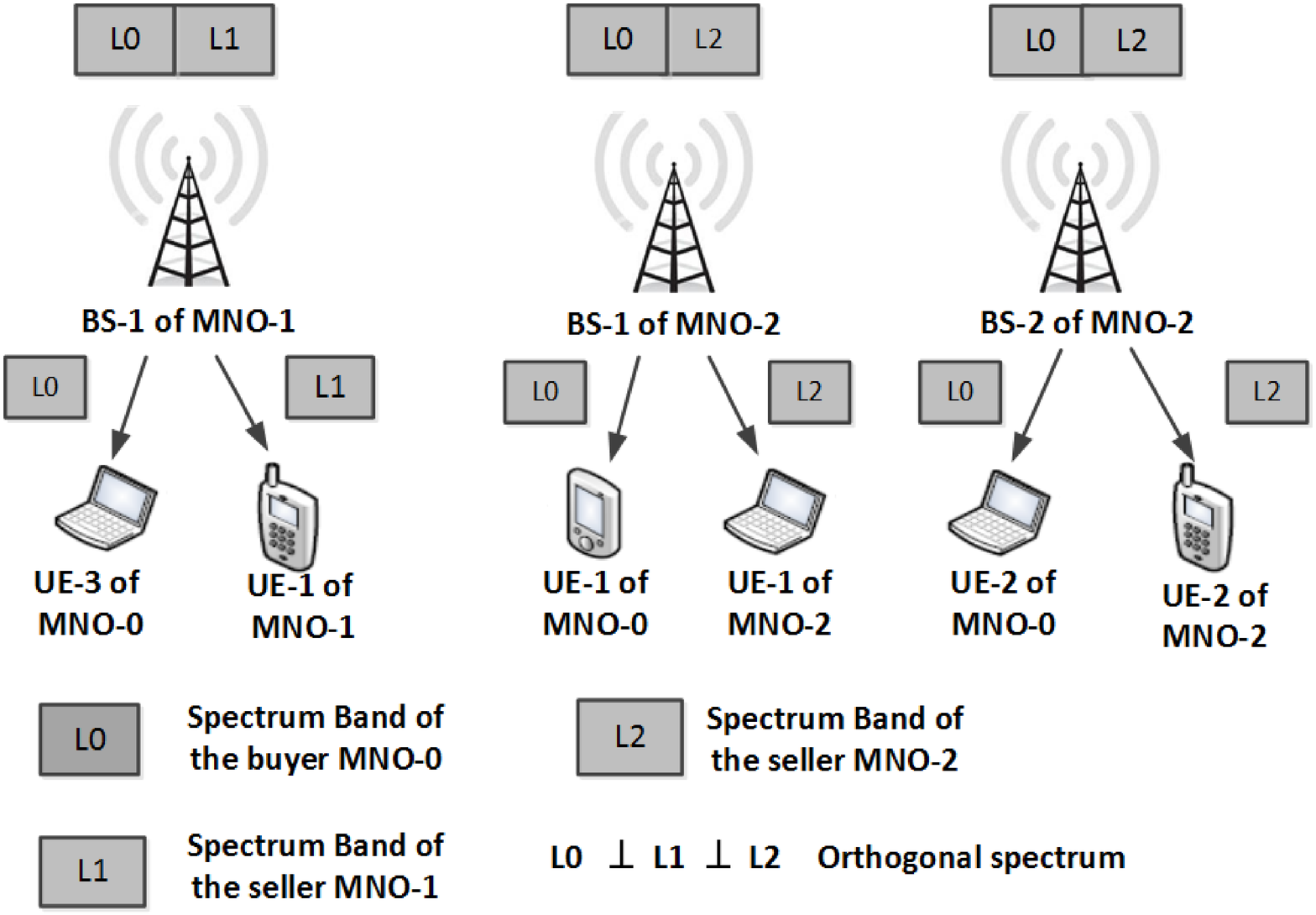}
        \caption{}
    \end{subfigure}%
    ~~
    \begin{subfigure}[h]{0.5\textwidth}
        \centering
        \includegraphics[height=2.1in]{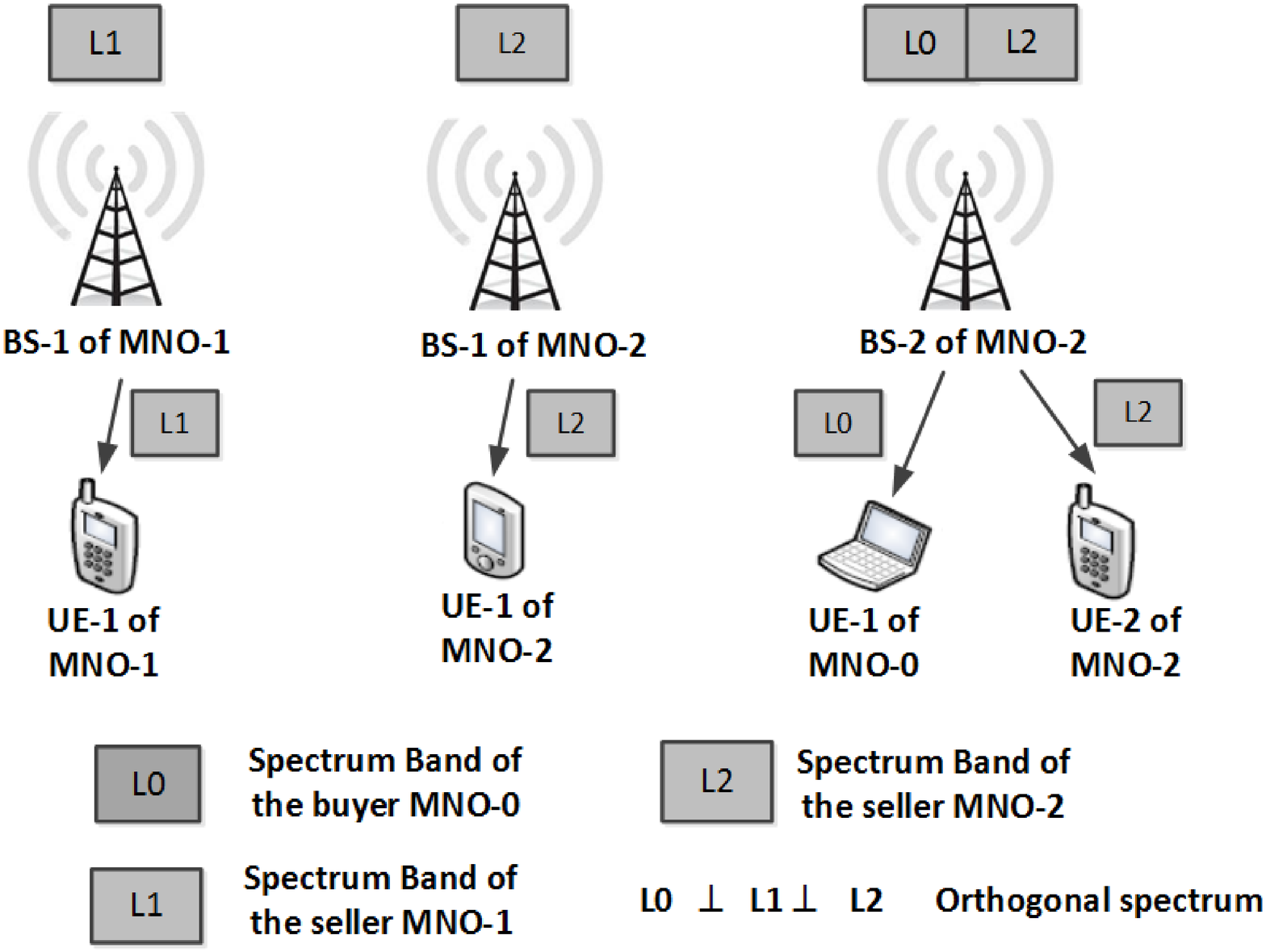}
        \caption{}
    \end{subfigure}
    \caption{The buyer MNO-$0$ buys infrastructure from seller MNOs.}
\label{fig:Infrasharing}
\end{figure}

Fig.~\ref{fig:Infrasharing} illustrates the scenarios when the MNO-$0$ buys infrastructure from two seller MNOs, namely, MNO-1 (with 1 BS) and MNO-2 (with 2 BSs). In Fig. 1(a), all of the BSs of MNO-1 and MNO-2 serve users (e.g. user equipment; UE-1, UE-2, and UE-3) subscribing to MNO-0 (which is described by \tbf{Assumption 2.1}), while in Fig. 1(b), only some shared BSs of seller MNOs (e.g. BS-2 of MNO-2) serve at least one UE of MNO-$0$ (which is described by \tbf{Assumption 2.2}).

\section{Analysis of Downlink SINR Coverage Probability for the Infrastructure Sharing System}
\label{sec:stogeoana}
Without loss of generality, we consider a typical UE of MNO-$0$ located at the origin, which associates with the nearest BS in the enlarged set of BSs given by $\Fcal$. We will denote the nearest BS from $\Fcal$ to the typical UE as BS-$0$.  We assume that the message signal undergoes Rayleigh fading with the channel power gain given by $g_0$. Furthermore, let $\alpha > 2$ denote the path-loss exponent for the path-loss model $r_0^{-\alpha}$, where $r_0$ is the distance between the typical UE and BS-$0$. Finally, let $\sigma^2$ denote the noise variance, and $p$ denote the transmit power of all the BSs in MNO-$0$, including BS-$0$. The downlink ${\rm SINR}$ at the typical UE is
${\rm SINR} = \frac{g_0 r_0^{-\alpha}p}{I + \sigma^2}$, where
$I$ is the interference experienced by a typical UE from the BSs that operate on the spectrum of MNO-$0$. These are the BSs that belong to MNO-$k$, where $k \in \Ncal \cup \{0\}$. Thus, $I = \sum_{i\in\Fcal \backslash \{0\}} \xi_i g_i r_i^{-\alpha}p$. Here $g_i$ is the co-channel gain between the typical UE and interfering BS-$i$, and $r_i$ is the distance between the typical UE and the interfering BS-$i$, where  $i\in\Fcal \backslash \{0\}$. The transmit power of each BS is $0 < p \leq p_{\max}$. Lastly, $\xi_i \in \{0,1\}$ is a binary variable indicating whether the BS-$i$ is active (if $\xi_i = 1$) or inactive (if $\xi_i = 0$) in the spectrum of MNO-$0$.


For a given threshold $T$, the SINR coverage probability for the typical UE of MNO-$0$'s cellular network is defined as:
$P_c = \mathrm{Pr}({\rm SINR} > T).$
While the case when $\lambda_A = \lambda_I$ has been explored in \cite{Andrews2011}, similar method can be used to find a more general formula for the coverage probability when $\lambda_A \neq \lambda_I$. Following \cite[Theorem 1]{Andrews2011}, we first condition on the nearest BS at the distance $r_0$ from a typical UE. The coverage probability averaged over $r_0$ is
\beq
P_c = \int_{r >0}\mathrm{Pr}({\rm SINR} > T \mid r_0)f_{r_0}(r)dr,
\eeq
where the probability density function of $r_0$ is given by, $f_{r_0}(r) = e^{-\pi \lambda_A r_0^2} 2 \pi \lambda_A r_0$. We have
\begin{align}
P_c &= \int_{r>0}\mathrm{Pr}\left( \frac{g_0 r_0^{-\alpha} p}{I + \sigma^2} > T \mid r_0 \right) e^{- \pi \lambda_A r^2} 2 \pi \lambda_A  r_0 d r_0.
\end{align}
Using the fact that the distribution of the Rayleigh fading channel power gain follows an exponential distribution $\exp\left(\frac{1}{p}\right)$, the coverage probability can be expressed as
\begin{align*}
\mathrm{Pr}[g_0 > T r_0^\alpha (\sigma^2 + I) \mid r_0] &= \mathbb{E}_I [e^{- \frac{T r_0^{\alpha}}{p} (\sigma^2 + I) }\mid r_0 ] = e^{- \frac{T \sigma^2 r_0^\alpha}{p}} \mathbb{L}_I \left(\frac{T r_0^{\alpha}}{p} \right),
\end{align*}
where $\mathbb{E}_I[.]$ is expectation taken with respect to the interference power, $\mathbb{L}_I(s)$ is the Laplace transform of the random variable $I$ evaluated at $s = \frac{T r_0^{\alpha}}{p}$, conditioned on the distance to the closest BS from the origin. It yields the coverage expression
\beq
P_c = \int_{r>0}e^{-\pi \lambda_A r_0^2} e^{-\frac{T \sigma^2 r_0^{\alpha}}{p}}\mathbb{L}_I\left(\frac{T r_0^{\alpha}}{p}\right) 2 \pi \lambda_A r_0 dr_0.
\label{eqn:coverage_laplace}
\eeq

For a homogeneous PPP,  $\mathbb{L}_I(\frac{T r_0^{\alpha}}{p})$ is given by
\beq
\mathbb{L}_I\left(\frac{T r_0^{\alpha}}{p}\right) = \exp \left\{ \lambda_I \pi (1 - \beta) r_0^2 \right\}, \quad \mbox{where}
\label{eqn:laplace}
\eeq
\beq
\beta = \frac{2 (T/p)^{2/\alpha}}{\alpha} \mathbb{E}_{g}[g^{2/\alpha} (\Gamma(-2/\alpha, T g/p))- \Gamma(-2/ \alpha)],
\label{eqn:beta}
\eeq
in which $\Gamma(z)$ is the Gamma function, while $\Gamma(z,a) = \int_z^{\infty} x^{a-1}e^{-x} dx$ is the upper incomplete Gamma function, and  $\mathbb{E}_g[.]$ is the expectation taken with respect to interferers' channel distribution $g$.
\begin{proposition}
The general expression of the coverage probability for the typical UE of MNO-$0$ cellular network is
\beq
P_c = \pi \lambda_A \int_0^{\infty}\exp \{-(Az + Bz^{\alpha /2})\} dz,
\label{eqn:coverage-integral}
\eeq
where the coefficients $A$ and $B$ are given by
\begin{align}
A = \pi[(\lambda_I(\beta -1))+ \lambda_A], \qquad B = \frac{T \sigma^2}{p}.
\label{coeff:AB}
\end{align}
\label{prop:coverage-general}
\end{proposition}

\begin{IEEEproof}
A formula for a coverage probability of the typical UE when the BSs are distributed according to a homogeneous PPP of intensity $\lambda$ is derived in \cite[Eqn.2]{Andrews2011}. Substituting (\ref{eqn:laplace}) in (\ref{eqn:coverage_laplace}) and changing the variable $r_0^2 \to z$, we can express the coverage probability as
\beq
P_c = \int_{z > 0} \underbrace{e^{- \frac{T \sigma^2}{p} z^{\alpha/2} }}_{\text{noise}} \underbrace{e^{-\pi(\lambda_I(\beta -1))z}}_{\text{interference}} \underbrace{e^{- \lambda_A \pi z} \pi \lambda_A}_{\text{user association}} dz.
\label{eqn:coverage_general}
\eeq
Here the integrand in (\ref{eqn:coverage_general}) comprises of terms related to  noise, interference, and user association, while each BS employs a constant power $p$. We can express (\ref{eqn:coverage_general}) as (\ref{eqn:coverage-integral}) with the coefficients $A$ and $B$ as given in (\ref{coeff:AB}). 
\end{IEEEproof} 
When the interfering links undergo Rayleigh fading, $\beta = 1 + \rho(T,\alpha)$, where
\beq
\rho(T,\alpha) = T^{2/\alpha} \int_{T^{-2/\alpha}}^\infty (1+u^{\alpha/2})^{-1} \ud u.
\label{eqn:rho}
\eeq
For this special case, we see that $\beta$ is independent of transmit power.

Except for $\alpha = 4$, $P_c$ cannot be evaluated in closed form. Nevertheless, a simple closed-form approximation for the general case, where $\alpha > 2$, and where both noise and intra-operator interference are present, can be given as \cite[Eqn. 4]{Sudarshan2016}
\begin{align}
P_c \simeq \pi \lambda_A \left[ A+ \frac{\alpha}{2} \frac{B^{2/\alpha}}{\Gamma\big(\frac{2}{\alpha}\big)} \right]^{-1},
\label{eqn:coverage-approx}
\end{align}
where $\Gamma(z)$ is the Gamma function. For ``interference-limited case'', which occurs when $\sigma^2 \to 0$, we have $B \to 0$; therefore, the above approximation simplifies to
\beq
P_c \simeq \frac{\lambda_A}{(\lambda_I(\beta -1))+ \lambda_A}.
\eeq

The significance of the approximation in (\ref{eqn:coverage-approx}) is that it allows us to study the asymptotic behavior of $P_c$. These asymptotic results give us a qualitative understanding of the system as various parameters change. Likewise, we can use (\ref{eqn:coverage-approx}) to obtain the required transmit power $p$ for given QoS, as we will see in the later sections.
%

In \textbf{Propositions 2} -- \textbf{5} in Sections \ref{Pc:Assump1and2.1} and \ref{Pc:Assump1and2.2} below, we examine the coverage probability of buyer MNO-$0$ when every BS (\textbf{Assumption 2.1}) and when some of BSs (\textbf{Assumption 2.2}) of seller MNOs serve the users of MNO-$0$. In \textbf{Proposition 6} in Section \ref{Pc:own}, we show the coverage probability for the case when MNO-0 uses its own infrastructure. Also, for all these cases, the asymptotic behavior of coverage probability is expressed accordingly.

\subsection{SINR Coverage Analysis When Assumption 2.1 Holds}
\label{Pc:Assump1and2.1}
\begin{proposition}
Under Assumption $1$ and Assumption $2.1$, the coverage probability of a typical UE of buyer MNO-$0$ is
\beq
P_c = \pi \lambda \int_0^{\infty}\exp \{-(A_1 z + Bz^{\alpha /2})\} dz,
\label{eqn:coverage-case1}
\eeq
where $A_1 = \pi \lambda \beta$, and $\beta$ and $B$ are given by (\ref{eqn:beta}) and (\ref{eqn:coverage-integral}), respectively. Then, we can approximate $P_c$ using (\ref{eqn:coverage-approx}) as
\beq
P_c \simeq \left[ \beta + \frac{\alpha}{2 \pi \lambda}\frac{B^{2/\alpha}}{\Gamma\big(\frac{2}{\alpha}\big)} \right]^{-1}.
\label{eqn:Pc_approx_case1}
\eeq
\end{proposition}

\begin{IEEEproof}
As per Assumption 2.1, we have $\lambda_I = \lambda_A = \lambda$. Substituting these values in Proposition \ref{prop:coverage-general}, we obtain (\ref{eqn:coverage-case1}). The approximation (\ref{eqn:Pc_approx_case1}) is obtained by applying (\ref{eqn:coverage-approx}).
\end{IEEEproof}

\begin{proposition}
Asymptotic behavior of $P_c$:
 (i) When the number of seller MNOs ($N$) is fixed, as the BS intensity of MNO-$0$ increases such that $\lambda_0 \to \infty$, the coverage of MNO-$0$ saturates at $\frac{1}{\beta}$. (ii) For a fixed BS intensity of MNO-$0$, as the number of seller MNOs ($N$) increases, the coverage probability also saturates at $\frac{1}{\beta}$. (iii) For fixed number of seller MNOs ($N$), if  MNO-$0$ does not have its own infrastructure such that $\lambda_0 = 0$, then $P_c \simeq \left[\beta+ \frac{\alpha}{2\pi \Gamma(\frac{2}{\alpha})} \frac{B^{2/\alpha}}{\sum_{i=1}^N \lambda_i}\right]^{-1}$.
\label{prop:limPc_case1}
\end{proposition}

\begin{IEEEproof}
From the closed-form approximation in (\ref{eqn:Pc_approx_case1}), we can see that (i) When $\lambda_0 \to \infty$, since $B$ and $\sum_{k=0}^N \lambda_k$ remain constant, $P_c \to 1/\beta$. (ii) When $N \to \infty$, since $B$ and and $\lambda_0 \beta$ are constants, $P_c \to 1/\beta$. (iii) When $\lambda_0 = 0$, $A_1 = \pi \beta \sum_{i=1}^N \lambda_i$. Simplifying (\ref{eqn:Pc_approx_case1}), we obtain the desired results.
\end{IEEEproof}

\subsection{SINR Coverage Analysis When Assumption 2.2 Holds}
\label{Pc:Assump1and2.2}

\begin{proposition}
Under Assumption 1 and Assumption 2.2, the downlink coverage probability of a typical user of buyer MNO-$0$ is
\beq
P_c =  \pi \lambda \int_0^{\infty}\exp \{-(A_2 z + Bz^{\alpha /2})\} dz,
\label{eqn:coverage-case2}
\eeq
where $A_2 = \pi ( \sum_{k \in \Ncal \cup \{0\}} w_k \lambda_k (\beta -1) + \lambda)$, and the $B$ and $\beta$ are given by (\ref{eqn:beta}) and (\ref{eqn:coverage-integral}), respectively.  Also, the approximate $P_c$ for this case is
 \begin{align}
P_c \simeq \left[ 1 + \frac{\bar{\lambda}(\beta -1)}{\lambda}  +  \frac{\alpha}{2\pi\lambda}\frac{B^{2/\alpha}}{\Gamma(\frac{2}{\alpha})} \right]^{-1},
\label{eqn:Pc_approx_case2}
\end{align}
where $\bar{\lambda} = \sum_{k \in \Ncal \cup \{0\}} w_k \lambda_k$ and $w_k = \lambda_k/\lambda$.
\end{proposition}

\begin{IEEEproof}
As per \textbf{Assumption 1}, we have $\lambda_A = \lambda$, while according to \textbf{Assumption 2.2}, we have $\lambda_I = \sum_{k \in \Ncal \cup \{0\}} w_k \lambda_k$. Substituting these values in \textbf{Proposition \ref{prop:coverage-general}}, we have the desired result in (\ref{eqn:coverage-case2}). Also, using the approximation (\ref{eqn:coverage-approx}) for (\ref{eqn:coverage-case2}), we obtain (\ref{eqn:Pc_approx_case2}).  Lastly, $w_k$ is the activity level of users of MNO-$0$ in another MNO-$k$, which is equivalent to the probability that a user associates with shared BSs belonging to MNO-$k$. That is, $w_k$ is the probability that the BS belonging to MNO-$k$ is the nearest BS to the typical user. Since the total intensity of BSs that a  user can associate with is $\lambda$, due to superposition property, $w_k = \lambda_k/\lambda$ is the probability that a user will connect to a BS belonging to MNO-$k$.
\end{IEEEproof}

\begin{proposition}
Asymptotic behavior of $P_c$:
 (i) For fixed number of seller MNOs ($N$), when the BS intensity of the buyer MNO-$0$ increases ($\lambda_0 \to \infty$), the coverage of MNO-$0$ saturates at $1/\beta$. (ii) For fixed BS intensity of MNO-$0$, as the number of seller MNOs ($N$) increases, the coverage of MNO-$0$ increases and then saturates to $1$. That is, if $\lim_{N\rightarrow \infty} \sum_{i\in\Ncal} \lambda_i = \infty$ and $\lim_{N\rightarrow\infty} \bar{\lambda}/\lambda = 0$,
then $\lim_{N\rightarrow \infty}P_c = 1$, (iii) For fixed number of seller MNOs ($N$), if  MNO-$0$ does not have its own infrastructure, then $P_c \simeq \left[1 + \frac{\bar{\lambda}(\beta -1)}{\sum_{i=1}^N \lambda_i} + \frac{\alpha}{2\pi \Gamma(\frac{2}{\alpha})} \frac{B^{2/\alpha}}{\sum_{i=1}^N \lambda_i}\right]^{-1}$.
\label{prop:limPc_case2}
\end{proposition}

\begin{IEEEproof}
From the approximation in (\ref{eqn:Pc_approx_case2}),  (i) using $w_i = \lambda_i/\lambda$, we have
\beq
\frac{\bar{\lambda}}{\lambda} = \frac{\sum_{i \in \Ncal \cup \{0\}} \lambda_i^2}{\lambda^2} =  \frac{\sum_{i \in \Ncal \cup \{0\}} \lambda_i^2}{(\sum_{i \in \Ncal \cup \{0\}} \lambda_i)^2}. \nonumber
\eeq
Dividing both the numerator and denominator on the right side by $\lambda_0^2$, we obtain
\[ \frac{\bar{\lambda}}{\lambda} = \frac{1 + \sum_{i \in \Ncal \cup \{0\}} (\frac{\lambda_i}{\lambda_0})^2}{[1 + \sum_{i \in \Ncal \cup \{0\}}  (\frac{\lambda_i}{\lambda_0})]^2}.\]
Therefore, when $\lambda_0 \rightarrow \infty$, we have $\lim_{\lambda_0 \rightarrow \infty}\frac{\bar{\lambda}}{\lambda} = 1$. Since $B$ remains constant, $\frac{\alpha}{2\pi\lambda}\frac{B^{2/\alpha}}{\Gamma(\frac{2}{\alpha})} \rightarrow 0$. Thus, we have $P_c \rightarrow 1/\beta$.

(ii) As $N\rightarrow\infty$, given our assumptions, $\bar{\lambda}/\lambda \rightarrow 0$. Similarly, since $B$ remains constant, $\frac{\alpha}{2\pi\lambda}\frac{B^{2/\alpha}}{\Gamma(\frac{2}{\alpha})} \rightarrow 0$. Thus, we have $P_c \rightarrow 1$.

(iii) When $\lambda_0 = 0$, we have $\lambda = \sum_{i \in \Ncal \cup \{0\}} \lambda_i$. Thus, we obtain the desired result.
\end{IEEEproof}

Note that \tbf{Proposition~\ref{prop:limPc_case2} (i)} is valid not only when $\lambda_0 \rightarrow \infty$, but also when any $\lambda_i \rightarrow \infty$. In our case, an increase in the BS intensity does not correspond to an increase in co-channel interference, which is different from \cite{Andrews2011}. \tbf{Proposition \ref{prop:limPc_case2}} also confirms our intuition that a greater sharing of infrastructure leads to a better coverage. Increasing the buyer MNO-$0$'s infrastructure also leads to coverage improvement.

We can see that when the number of seller MNOs ($N$) increases, the coverage of the buyer MNO-$0$ saturates at $1$ following  \tbf{Proposition~\ref{prop:limPc_case2} (ii)}. This leads to the main difference between  \tbf{Proposition~\ref{prop:limPc_case1}} and \tbf{Proposition~\ref{prop:limPc_case2}}. In \tbf{Proposition~\ref{prop:limPc_case1}}, although the number of sellers increased, the maximum coverage of MNO-$0$ leads to only $1/\beta$.

\subsection{When the MNO-$0$ Employs its Own Infrastructure}
\label{Pc:own}

For the case when the MNO-$0$ uses its own infrastructure, the coverage probability can be obtained as follows.
\begin{proposition}
\label{prop:Pc_noinfra}
The coverage of MNO-$0$ without buying infrastructure is approximated as
\beq
P_c \simeq \left[ \beta + \frac{\alpha}{2 \pi \lambda_0}\frac{B^{2/\alpha}}{\Gamma\big(\frac{2}{\alpha}\big)} \right]^{-1}.
\label{eqn:Pc_noinfra}
\eeq
\end{proposition}

\begin{IEEEproof}
Without infrastructure sharing, the interference is only from BSs of MNO-$0$. Thus, $\lambda_I = \lambda_0$, regardless of \textbf{Assumption 2.1} or \textbf{Assumption 2.2}. Also, the user can associate only with the BSs of MNO-$0$. Thus, $\lambda_A = \lambda_0$. Hence, using (\ref{eqn:coverage-approx}) we obtain the desired result.
\end{IEEEproof}

Asymptotically, $P_c \to 1/\beta$ as $\lambda_0 \to \infty$, since $B$ remains constant.

\section{Analysis of Trade-offs}
\label{subsec:Tx_infra}

\subsection{Minimum Transmit Power Required to Satisfy the QoS}
\label{subsec:Tx_covQoS}
In the \textbf{Propositions 7} and \textbf{8} below, we express the minimum transmit power for each BS of the buyer MNO-$0$ to achieve the coverage QoS when every BS (\textbf{Assumption 2.1}) and when some of BSs (\textbf{Assumption 2.2})  of seller MNOs serve the users of MNO-$0$. Let us further assume that the MNO-$0$ wants to ensure that the coverage probability of a typical UE satisfies the QoS constraint
\beq
P_c \geq 1 - \epsilon,
\label{eqn:QoS-constrain}
\eeq
where $0< \epsilon < 1$ is some arbitrary value. In order to satisfy the coverage constrain in (\ref{eqn:QoS-constrain}), the minimum power required for each BS of MNO-$0$, for given infrastructure, is given by the following proposition.

\begin{proposition}
Let Assumption 1 and Assumption 2.1 hold. Assume that the interfering links undergo Rayleigh fading and $\lambda$ be defined as before. Then, given the  condition that $1- \epsilon < 1/\beta$, where $\beta$ is given by (\ref{eqn:beta}), the minimum transmit power required for each BS of MNO-$0$ such that $P_c \geq 1 - \epsilon$, is
\beq
p \simeq c_0 \lambda^{-\alpha/2}, \quad \mbox{where } \quad  c_0 = \left[ \frac{ 2 \pi (1-(1-\epsilon)\beta)}{\alpha (1-\epsilon)(T \sigma^2 )^{2/\alpha}} \Gamma \left(\frac{2}{\alpha}\right) \right]^{-\alpha / 2}.
\label{eqn:min-power-case1}
\eeq
\label{prop:power-infra-trade-off-case1}
\end{proposition}

\begin{IEEEproof}
When the interfering links undergo Rayleigh fading, $\beta = 1 + \rho$, as given in (\ref{eqn:rho}), and is independent of $p$. Thus, using (\ref{eqn:coverage-approx}) in the inequality $P_c \geq 1 - \epsilon$, and solving for $p$, we obtain the desired result. For $p > 0$, it suffices that $1 - \beta (1- \epsilon) > 0$ in the expression for $c_0$. Re-arranging the terms gives the desired result.
\end{IEEEproof}

\begin{proposition}
Let Assumption 1 and  Assumption 2.2 hold. We assume that the interfering links undergo Rayleigh fading and $\lambda$ be defined as before. Then, given the sufficient condition that $1-\epsilon < 1/\beta'$, where $\beta' = 1 + \frac{\bar{\lambda}(\beta -1)}{\lambda}$, the minimum transmit power required for each BS of MNO-$0$ such that $P_c \geq 1 - \epsilon$, is
\beq
p \simeq \widehat{c}_0 \lambda^{-\alpha / 2}, \quad \mbox{where } \quad \widehat{c}_0 = \left[ \frac{ 2 \pi (1-(1-\epsilon)\beta')}{\alpha (1-\epsilon)(T \sigma^2 )^{2/\alpha}} \Gamma(\frac{2}{\alpha}) \right]^{-\alpha / 2}.
\label{eqn:min-power-case2}
\eeq
\label{prop:power-infra-trade-off-case2}
\end{proposition}

\begin{IEEEproof}
When the interfering links undergo Rayleigh fading, $\beta = 1 + \rho$, as given in (\ref{eqn:rho}), and is independent of $p$. Thus, using (\ref{eqn:coverage-approx}) in the inequality $P_c \geq 1 - \epsilon$, and solving for $p$, we obtain the desired result. For $p > 0$, it suffices that $1-(1-\epsilon)(1+\beta') > 0$ in the expression for $\widehat{c}_0$. Here $\beta' = 1 + \frac{\bar{\lambda}(\beta -1)}{\lambda}$ and $\bar{\lambda} = \sum_{i=0}^N w_i \lambda_i$.
Re-arranging the terms gives the sufficient condition.
\end{IEEEproof}


According to \tbf{Proposition~\ref{prop:power-infra-trade-off-case1}} and \tbf{Proposition~\ref{prop:power-infra-trade-off-case2}}, for a given QoS, the transmit power of the BSs of an MNO should decrease with increasing BS intensity.
For instance, if $\alpha = 4$, then $p \propto \frac{1}{\sqrt{\lambda}}$. Then, we will use (\ref{eqn:min-power-case1}) for \tbf{Assumption 2.1}, and (\ref{eqn:min-power-case2}) for \tbf{Assumption 2.2} to obtain the minimum transmit power from BS of MNO-$0$, when there is no infrastructure sharing. Also, we will obtain the optimal cell radius of a BS of MNO-$0$ when using the minimum power.

\subsection{Trade-off Between Power and Infrastructure}
\label{subsec:useroutage}
Every MNO wishes to guarantee a certain probability of coverage to its own customers. For this purpose, if a UE is experiencing outage, the MNO can either choose to increase the transmit power of the BSs so as to increase the coverage radius, or offload the call to a shared BS. It is natural to look at the possible trade-off between increasing the power and sharing more infrastructure. 


Intuitively, in both \tbf{Proposition~\ref{prop:power-infra-trade-off-case1}} and \tbf{Proposition~\ref{prop:power-infra-trade-off-case2}}, the minimum required transmit power decreases with increasing BS densification.
\begin{itemize}
\item For \tbf{Assumption 2.1} the sufficiency condition $1/\beta$ is the maximum attainable coverage probability as the transmit power $p \rightarrow \infty$ and as the system becomes interference limited (i.e., $B \rightarrow 0$).
That is, we have the upper bound $P_c \leq 1/\beta$. Thus, the QoS, $1 - \epsilon$, can be achieved by varying the transmit power only when $1- \epsilon < 1/\beta$.
\item Similarly, for \tbf{Assumption 2.2}, the sufficiency condition $1/\beta'$ is the maximum attainable coverage probability when the transmit power $p \rightarrow \infty$ such that $B \rightarrow 0$. The QoS, $1-\epsilon$, can be achieved by varying the transmit power only when $1- \epsilon < 1/\beta'$.
\end{itemize}

We can see that if the sufficient conditions for the cases with \tbf{Assumption 2.1} and  \tbf{Assumption 2.2} are violated, MNO-$0$ cannot satisfy the outage QoS by simply varying the transmit power. The MNO-$0$ will have to buy more infrastructure from other MNOs.

%
Let $R$ be the cell radius of a BS defined as the distance at which a UE will receive $-3$ dB SNR. Then, for the important special cases when there is no infrastructure sharing,  we have the following scaling law as a corollary.

\begin{corollary}
When there is no infrastructure sharing, the minimum BS transmit power of \tbf{Assumption 2.1} and \tbf{Assumption 2.2} for which $1- \epsilon < 1/\beta$ are
\begin{enumerate}
\item From \tbf{Assumption 2.1}, the minimum transmit power is $p \simeq c_0 \lambda_0^{-\alpha/2}$, where $c_0$ is given in (\ref{eqn:min-power-case1}) and $c_0$ is independent of $\lambda_0$. The optimal cell radius is $R \simeq \frac{c_0'}{\sqrt{\lambda_0}}$, where $c_0' = \left( \frac{2 c_0}{\sigma^2} \right)^{1/\alpha}$.
\item From \tbf{Assumption 2.2}, the minimum transmit power is $p \simeq \widehat{c}_0 \lambda_0^{-\alpha/2}$, where $\widehat{c}_0$ is given in (\ref{eqn:min-power-case2}) such that $\beta' = \beta$, and $\widehat{c}_0$ is independent of $\lambda_0$. We can obtain the optimal cell radius as $R \simeq \frac{\widehat{c}_0'}{\sqrt{\lambda_0}}$, where $\widehat{c}_0' = \left( \frac{2 \widehat{c}_0}{\sigma^2} \right)^{1/\alpha}$.
\end{enumerate}
\label{coro:power-infra-trade-off}
\end{corollary}
\begin{IEEEproof}
When there is no infrastructure sharing, it means that $\Ncal = \emptyset$ and we have $\lambda = \lambda_0$. Since the cell edge is defined as the distance at which SNR is $-3$ dB, we have
\beq
\frac{p R^{-\alpha}}{\sigma^2} = \frac{1}{2}.
\label{eqn:cell_edge_dist}
\eeq
Using \tbf{Proposition~\ref{prop:power-infra-trade-off-case1}} and \tbf{Proposition~\ref{prop:power-infra-trade-off-case2}}, the proofs are as follows:
(i) Given \tbf{Assumption 2.1}, putting $p = c_0 \lambda_0^{-\alpha/2}$ from (\ref{eqn:min-power-case1}) in (\ref{eqn:cell_edge_dist}), we can solve for $R$ to obtain the result.
(ii) Similarly, given \tbf{Assumption 2.2}, in (\ref{eqn:min-power-case2}) $\bar{\lambda} = \lambda_0$ without infrastructure sharing. Also $\beta' = \beta$. Thus, we obtain required minimum transmit power in the Corollary \ref{coro:power-infra-trade-off}. Then, putting $p = \widehat{c}_0 \lambda_0^{-\alpha/2}$ in (\ref{eqn:cell_edge_dist}), we can solve for $R$ to obtain the result.

\end{IEEEproof}
A scaling law similar to \tbf{Corollary \ref{coro:power-infra-trade-off}} can be found in \cite[Lemma 1]{Sarkar2014} and \cite[Lemma 1]{Perabathini2014} for homogeneous PPP, using a slightly different approximation, as $p \propto \lambda_0^{-\alpha/2+1}$. However, our formula differs from theirs in the order of the exponent as well as the proportionality constant.  Likewise, the scaling law for the  cell radius, $R \propto 1/\sqrt{\lambda_0}$, corresponds to that obtained by \cite{Hanly2002} for hexagonal grid model.

\subsection{Areal Power Consumption of Seller MNO}
\label{Arealpower:BSinten}

In \textbf{Propositions 9} and \textbf{10} below, we express the areal power consumption  as a function of BS intensity of seller MNO and show the convexsity of the areal power consumption.

Let the transmit power of each BS belonging to the seller MNO-$k$, where $k \in \Kcal \backslash \{0\}$, be denoted by $p_k$. Apart from the transmit power, each BS also consumes a fixed amount of circuit power, denoted by $p_c$. Hence, the total power consumption of a  BS of an MNO-$k$ is $p_k + p_c$. Since the MNO-$k$ has $\lambda_k$ BS per unit area, the areal power consumption of the network (i.e., power consumption per unit area) is
\beq
S_k = \lambda_k ( p_k + p_c).
\label{eqn:power-per-area}
\eeq

For MNO-$k$, let the QoS constaint on coverage probability of a typical UE be $P_c \geq 1 - \epsilon$ and the threshold SINR be $T_k$. In order to satisfy this constraint, it can either increase its BS intensity $\lambda_k$ or increase its transmit power $p_k$. The trade-off between $\lambda_k$ and $p_k$ was given by \tbf{Proposition \ref{prop:power-infra-trade-off-case1}}.
Similarly, the trade-off between $\lambda_k$ and $S_k$ follow immediately.

\begin{proposition}
Given the assumptions in \tbf{Proposition~\ref{prop:power-infra-trade-off-case1}}, the areal power consumption of seller MNO-$k$, where $k \in \Kcal \backslash \{0\}$, is
\beq
S_k(\lambda_k) = \left\{ \begin{array}{ll}
\lambda_k(p_{\max} + p_c), & \mathrm{if} \quad  0 \leq \lambda_k \leq (\frac{c_k}{p_{\max}})^{2/\alpha}, \\
\lambda_k(c_k \lambda_k^{-\alpha/2} + p_c), & \mathrm{if} \quad  \lambda_k \geq (\frac{c_k}{p_{\max}})^{2/\alpha},
\end{array} \right.
\label{eqn:areal-power}
\eeq
where $c_k = \left[\frac{2 \pi ( 1- (1-\epsilon)\beta ) }{\alpha (1-\epsilon)( T_k \sigma^2)^{2/\alpha}} \Gamma(\frac{2}{\alpha}) \right]^{-\alpha / 2}$.
\end{proposition}
\begin{IEEEproof}
Since the MNO-$k$ does not buy infrastructure from other MNOs, the net BS intensity that a typical UE of MNO-$k$ experiences is $\lambda_k$. Thus, from \textbf{Corollary \ref{coro:power-infra-trade-off}}, we have $p_k \simeq c_k \lambda_k^{-\alpha/2}$. Putting $p_k$ in (\ref{eqn:power-per-area}) and recalling that $0 < p_k \leq p_{\max}$, we have (\ref{eqn:areal-power}).
\end{IEEEproof}

We see that $S_k$ is a piece-wise continuous function of $\lambda_k$ which initially increases linearly with $\lambda_k$, and beyond a certain point, it behaves as a convex function. This can be verified by checking the second derivative of $S_k$ for $\lambda_k \geq (\frac{c_k}{p_{\max}})^{2/\alpha}$ as
\beq
\frac{\ud^2 S_k}{\ud \lambda_k^2} = \frac{c_k \alpha (\alpha - 2)}{4} \lambda_k^{-\frac{\alpha}{2} - 1}.
\label{eqn:seconddiff_Sk}
\eeq
Since $c_k >0$ and $\alpha > 2$, we have $\frac{\ud^2 S_k}{\ud \lambda_k^2} > 0$, proving the convexity of $S_k$ in the region $\lambda_k \geq (\frac{c_k}{p_{\max}})^{2/\alpha}$. As such, studying the behaviour of $S_k$ is not straightforward.  Nevertheless, the local minima in the convex region can be found.

\begin{proposition}
Given the assumptions in \tbf{Proposition~\ref{prop:power-infra-trade-off-case1}~}, let $\lambda_{th} = \left(\frac{c_k}{p_{\max}}\right)^{2/\alpha}$. Then, for the region $\lambda_k \geq \lambda_{th}$, the BS intensity for which the areal power consumption of MNO-$k$, where $k \in \Kcal \backslash \{0\}$, is minimum is
\beq
\lambda_{k,\min} = \max\left(\lambda_{th},\left[\frac{c_k}{p_c}\left(\frac{\alpha}{2} -1 \right)\right]^{2/\alpha}\right).
\eeq
\end{proposition}
\begin{IEEEproof}
We have $\ud S_k/\ud \lambda_k = p_c - (c_k(\alpha - 2) \lambda_k^{-\alpha/2})/2$. Solving $\ud S_k/\ud \lambda_k = 0$ for $\lambda_k$, we have $\lambda_k^{*} = [\frac{c_k}{p_c}(\frac{\alpha}{2} -1)]^{2/\alpha}$. This is clearly the minima if $\lambda_{th} < \lambda_k^{*}$. Otherwise, $\lambda_{k,\min} = \lambda_{th}$.
\end{IEEEproof}

\begin{figure}[h]
\centering
\includegraphics[height=3.8 in, width=3.8 in, keepaspectratio = true]{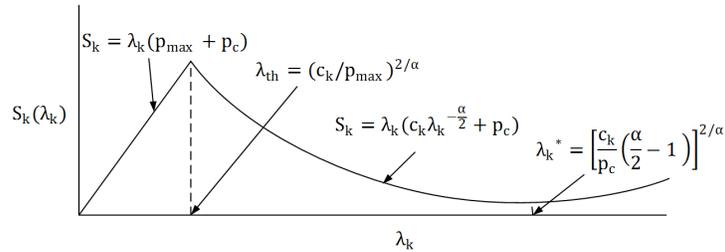}
\caption{The areal power consumption ($S_k$) versus BS intensity ($\lambda_k$).}
\label{fig:Sk_vs_lambdak}
\end{figure}

In Fig.~\ref{fig:Sk_vs_lambdak}, we illustrate $S_k$ as a function of $\lambda_k$, as given in (\ref{eqn:areal-power}). We can see that it is composed of linear and convex parts. The convex part of $S_k$ corresponds to that obtained for hexagonal grid models via simulations in \cite{Richter2009}. Similar, but not the same, formulas were given in \cite{Sarkar2014,Perabathini2014}.

\section{Analysis of Market: Buyer's Strategy}\label{section:strategybuyer}
In this section, we propose a strategy for the buyer MNO-$0$ which will allow it to choose the seller MNOs to buy the infrastructure from. By using our method, the buyer MNO will select the necessary number of seller MNOs, at minimum cost, such that it can serve its UEs guaranteeing some QoS. We have already seen in \tbf{Propositions \ref{prop:power-infra-trade-off-case1} and \ref{prop:power-infra-trade-off-case2}} and the discussion given in Section \ref{subsec:useroutage} that when the QoS constrain is such that $1 - \epsilon < 1/\beta'$ and $1-\epsilon < 1/\beta$, respectively, the QoS can be satisfied by simply increasing the transmit power. As such, in this section, we will consider the case when these conditions are violated. We first have the following proposition:

\begin{proposition}
Under Assumption 2.1, when the QoS condition $P_c \geq 1 - \epsilon > 1/\beta$, the MNO-$0$ cannot improve its coverage by buying infrastructure.
\end{proposition}
\begin{IEEEproof}
Using the approximation (\ref{eqn:Pc_approx_case1}) in (\ref{eqn:QoS-constrain}) and solving for $\lambda$, we obtain the feasible constraint of $\lambda$ as
\beq
\lambda \geq \frac{\theta (1-\epsilon)}{1 - \beta(1-\epsilon)}.
\label{eqn:netBS_intensity_case1}
\eeq
where $\theta = \frac{\alpha}{2\pi} \frac{B^{2/\alpha}}{\Gamma\big(\frac{2}{\alpha}\big)}$, for some $\Ncal$ such that $\emptyset \subseteq \Ncal \subseteq \Kcal$. For positivity, the denominator of (\ref{eqn:netBS_intensity_case1}) should be $1 - \beta(1-\epsilon) > 0$, which when re-arranged gives $1/\beta > 1 - \epsilon$. This contradicts our assumption that $1-\epsilon > 1/\beta$.
\end{IEEEproof}

Although this result seems counter-intuitive, it is not surprising when we recall that in \tbf{Proposition~\ref{prop:limPc_case1} (ii)}, even when the number of MNOs which are willing to share infrastructure increases i.e. $N \to \infty$, the coverage of the BSs of MNO-$0$ can only achieve $P_c \to 1/\beta$. Accordingly, it is not possible for the MNO-$0$ to achieve the coverage beyond $1/\beta$ by buying more infrastructure for the scenario of \tbf{Assumption 2.1}.

\begin{proposition}
Under Assumption 2.2, for the QoS condition $P_c \geq 1 - \epsilon > 1/\beta'$ to be feasible for the buyer MNO-$0$, the net BS intensity $\lambda = \lambda_0 + \sum_{i \in \Ncal} \lambda_i$ must satisfy
\beq
\sum_{i\in\Ncal\cup\{0\}} \left(1 - \frac{w_i (\beta-1)(1-\epsilon)}{\epsilon} \right)\lambda_i \geq \frac{\theta (1-\epsilon)}{\epsilon},
\eeq
where $w_i = \frac{\lambda_i}{\lambda}$ and $\theta$ is as given in (\ref{eqn:netBS_intensity_case1}), for some $\Ncal$ such that $\emptyset \subseteq \Ncal \subseteq \Kcal$.
\end{proposition}
\begin{IEEEproof}
Using the approximation (\ref{eqn:Pc_approx_case2}) in (\ref{eqn:QoS-constrain}) and substituting $\bar{\lambda}= \sum_{i=0}^N w_i \lambda_i$, and solving for $\lambda$, we have the required result.
\end{IEEEproof}

If there is a cost associated with the sharing of infrastructure, then we can formulate a cost minimization problem with the QoS constraint as follows:
\begin{align}
\label{eqn:mincost_case2}
\text{min} & \sum_{k \in \Kcal \backslash \{0\} } q_{k} x_{k} \\
\text{s.t.} \quad \mathrm{(C1)} & \sum_{k \in \Kcal } \left[ 1 - \frac{\lambda_k x_k (\beta -1)(1-\epsilon)}{\lambda \epsilon} \right] \lambda_{k} x_{k} \geq  \frac{\theta(1-\epsilon)}{\epsilon}, \nonumber
\end{align}
where $q_k>0$ is the price of infrastructure when buying from MNO-$k$, where $k \in \Kcal \backslash \{0\}$, $x_k$ ($0 \leq x_{k} \leq 1$) denotes the fraction of infrastructure bought from seller MNO-$k$. Also, note that $x_0 = 1$ since the buyer cannot buy infrastructure from itself. We can interpret $x_k$ in two possible ways: 1) The buyer MNO-$0$ buys the entire infrastructure of MNO-$k$ but utilizes the whole infrastructure of MNO-$k$ for only $x_k$ fraction of time, or 2) the MNO-$0$ buys only a fraction $x_k$ of the total infrastructure of MNO-$k$, but utilizes it all the time. Thus the total amount of infrastructure bought from MNO-$k$ is $\lambda_k x_k$. Since the objective function is linear, the single constraint is quadratic in $x_k$, and the variable $x_k$ is a fraction in $[0,1]$, the problem (\ref{eqn:mincost_case2}) becomes a convex optimization problem.

The solution to the problem can be split into two parts. First, we need to find the optimal set  $\Ncal$, i.e., select MNOs from which to buy the infrastructure. Second, we need to determine the optimal value of $x_k$ for $k\in \Ncal$. Assuming that we know $\Ncal$, we find the optimal value of $x_k$ using Lagrange multiplier method. We have the Lagrangian of (\ref{eqn:mincost_case2}) as
\begin{align}
L \left( \mu ,x_k \right) &= \sum_{k\in\Ncal\cup\{0\}} q_k x_k - \frac{1}{\mu}\bigg( \frac{\theta (1- \epsilon)}{\epsilon} - \sum_{k\in\Ncal\cup\{0\}} \lambda_k x_k  + \sum_{k\in\Ncal\cup\{0\}} \frac{\lambda_k^2 x_k^2}{\lambda} (\beta -1)(1- \epsilon)\bigg).
\end{align}
Here $1/\mu$ is the Lagrange multiplier such that $\mu > 0$. Note that here for $k=0$, $q_0 = 0$ and $x_0 = 1$. Taking first order partial derivative of $L(\mu, x_k)$ with respect to $x_k$ for $k\in\Ncal$, we obtain
\beq
\frac{\partial L(\mu, x_k)}{\partial x_k} = q_k + \frac{\lambda_k}{\mu} - \frac{2 \lambda_k^2 x_k (\beta-1)(1-\epsilon)}{\mu \lambda}.
\eeq
By using the first order optimality condition $\frac{\partial L}{\partial x_k} =0$, we obtain the optimal $x_k^{*}$ as
\beq
x_k^{*} = \frac{\lambda (\mu q_k + \lambda_k)}{2 \lambda_k^2 (\beta -1)(1- \epsilon)}.
\label{eqn:optimal_xk}
\eeq
Substituting $x_k^{*}$ from (\ref{eqn:optimal_xk}) in the (C1) of (\ref{eqn:mincost_case2}) and setting it to be an equality, as per complementary slackness condition, yields $\sum_{k\in\Ncal\cup\{0\}} \left( 1 - \frac{\lambda_k x_k^* (\beta-1)(1-\epsilon)}{\lambda \epsilon} \right) \lambda_k x_k^* = \frac{\theta (1-\epsilon)}{\epsilon}$. Then, solving for optimal $\mu^*$ gives,
\beq
\mu^{*} = \frac{- F \pm \sqrt{F^{2} - 4 E (G - H)}}{2 E}
\label{eqn:gen_mu}
\eeq
where
\begin{align*}
E &= \sum_{k\in\Ncal\cup\{0\}} \frac{\lambda q_k^2}{4 \lambda_k^2 \epsilon},  & F &= \sum_{k\in\Ncal\cup\{0\}} q_k \lambda \left( \frac{1}{2 \lambda_k \epsilon} - \frac{1}{2 \lambda_k^2} \right),  \\
G &= \sum_{k\in\Ncal\cup\{0\}} \lambda \left( \frac{1}{ 4 \epsilon} - \frac{1}{2 \lambda_k} \right), & H &= \frac{\theta (1-\epsilon)^2 (\beta-1)}{\epsilon}.
\end{align*}

For the uniqueness of $\mu^*$, we have an extra condition that the discriminant be zero, i.e, $F^2 - 4E (G - H) = 0$ has to be satisfied. As such, by substituting $E$ and $F$ back to  (\ref{eqn:gen_mu}), and setting $F^2 - 4E (G - H) = 0$, we obtain the unique solution of $\mu^*$ as,
\begin{align}
\mu^* = \frac{\sum_{k\in\Ncal} \frac{q_k}{\lambda_k} \big( \frac{\epsilon}{\lambda_k} - 1 \big)}{ \sum_{k\in\Ncal} \left(\frac{q_k}{\lambda_k}\right)^2}.
\label{eqn:optimal_mu}
\end{align}
Positivity of $\mu^*$ is ensured if $\epsilon > \lambda_k$ for all $k\in\Ncal$.

Finally, substituting the unique $\mu^*$ from (\ref{eqn:optimal_mu}) to (\ref{eqn:optimal_xk}), we obtain the optimal and unique solution of $x_k^*$. It is easy to verify that the obtained $x_k^*$ is positive, since in (\ref{eqn:optimal_xk}), the only way that $x_k^*$ can be negative is when the denominator is negative, i.e., $(\beta - 1)(1 - \epsilon) < 0$. This condition is equivalent to $\epsilon > 1$, which is a contradiction. Thus, $x_k$ is always positive. However, the computed value of $x_k^*$ can be greater than unity. Hence, we set $x_k^* = 1$ if $x_k^* > 1$.

Let us now consider finding the optimal $\Ncal$, which is essentially a combinatorial problem. A simple greedy approach can be used for the selection of MNOs in $\Ncal$ \cite[Chap 17.1]{Korte2012}. The greedy algorithm is provided in \textbf{Algorithm \ref{alg:GreedyAlgo}} to solve the problem in (\ref{eqn:mincost_case2}). The idea behind this greedy algorithm is as follows: We first sort the seller MNOs according to their cost per BS intensity values in an ascending order. We then select the first MNO in this list and compute $\mu^*$ and $x_k^*$ from (\ref{eqn:optimal_mu}) and (\ref{eqn:optimal_xk}), respectively. If $x_k > 1$, then set $x_k = 1$. After this, we check whether the constrain (C1) in (\ref{eqn:mincost_case2}) is satisfied. If the (C1) is not satisfied, then we  take the next MNO from the list and repeat the procedure until the constrain is satisfied. The computational complexity of \textbf{Algorithm \ref{alg:GreedyAlgo}} is $\Ocal(K^2)$, since we need to compute $x_k^*$ for all $k \in \Ncal$ in two nested loops. A solution can be derived within $K^2$ iterations.

\begin{algorithm}
\caption{Greedy Algorithm}
\label{alg:GreedyAlgo}
 \begin{algorithmic}[1]
 \STATE Initialize $x_k =0$, $w = 0$,
 \STATE Compute $\rho_k= q_k/\lambda_k$
 \STATE Sort the sellers by $\rho_k$ in ascending order such that $\rho_{\pi_1} \leq \rho_{\pi_2} \cdots \leq \rho_{\pi_K}$
 \FOR{$i = 1$ \TO $K$}
       \STATE Set $\Ncal = \{\pi_1,\ldots, \pi_i \}$
       \STATE Compute $\mu^*$ using (\ref{eqn:optimal_mu})
       \STATE Compute $x_k^*$ using (\ref{eqn:optimal_xk}) for all $k \in \Ncal$
       \IF{ $x_k^* > 1$ for any $k \in \Ncal$}
       	\STATE Set $x_k^* = 1$,
       \ENDIF
        \IF{$\sum_{k \in \Ncal\cup\{0\}} \left[ 1 - \frac{\lambda_k x_k^* (\beta -1)(1-\epsilon)}{\lambda \epsilon} \right] \lambda_{k} x_{k}^* \geq  \frac{\theta(1-\epsilon)}{\epsilon},$}
             \STATE Terminate
         \ENDIF
 \ENDFOR
 \STATE Compute $P_c$ from (\ref{eqn:Pc_approx_case2}).
 \end{algorithmic}
 \end{algorithm}

\section{Analysis of Market: Sellers' Competition} \label{section:cournotseller}

\subsection{Cournot-Nash Oligopoly Market Model}

In this part, we will study the equilibrium pricing due to the sellers' competition as well as the optimal fraction of infrastructure that the seller MNOs will be willing to sell. We will formulate the sellers' competition as a Cournot-Nash oligopoly game \cite{Friedman1983}. The Cournot oligopoly model is suitable when there are limited number of sellers competing to sell homogeneous products. In our case, the product is the infrastructure provided by the seller MNOs. In the real world scenario, there are limited number of MNOs that compete to sell some amount of their infrastructure. Each seller MNO is selfish and is always interested in getting a better payoff. The MNOs do not communicate with each other, thus they will not know exactly how much infrastructure is being sold by their competitors. Also, the pricing will depend on the operational expense of the shared infrastructure. As such, we will consider the cost of power consumption. Note that the cost of spectrum does not come into play, since each seller MNO utilizes the spectrum of the buyer MNO-$0$  to serve the users subscribing to the buyer MNO-$0$.

Let the fraction of infrastructure to be sold from the seller MNO-$k$, $k\in \Kcal \backslash \{0\}$, be $z_k$, where $0 \leq z_k \leq 1$. Then, the total amount of infrastructure sold by the seller MNO-$k$ is $y_k = \lambda_k z_k$. Let the cost of operating its infrastructure be $C_k(y_k)$, which we define as
\beq
C_k(y_k) = a_k S_k(y_k) + d_k,
\label{eqn:Ck}
\eeq
where $a_k$ is the price of areal power consumption, $d_k$ is a fixed operation cost, and $S_k$ is as given in (\ref{eqn:areal-power}).
Let the overall infrastructure from $K$ seller MNOs available in the market be denoted by $y = \sum_{k=1}^K y_k $. Also, let us denote the fraction of infrastructure of all MNOs except MNO-$k$ by $y_{-k} = y - y_{k}$. 

Let the selling price of the infrastructure be $Q(y)$. In accordance to the ``law of demand" of economics, the seller MNOs will reduce the price when there is higher demand for the infrastructure in the market. We will assume $Q(y)$ to be
\beq
Q(y) = \theta - \eta y,
\label{eqn:revenueQ}
\eeq
where $\theta>0$ is the initial installation price of infrastructure from all seller MNOs and $\eta >0$ denotes the marginal price of the total infrastructure $y$ in the market. The MNO-$k$'s profit is
\begin{align}
F_{k}(y_1,\ldots,y_k) = y_k Q(y) - C_k( y_k).
\label{eqn:profit}
\end{align}

In order to maximize the profit of MNO-$k$ with respect to $y_k$, we first partially differentiate (\ref{eqn:profit}) with respect to $y_k$, and noting that $\partial y/ \partial y_k = 1$, we obtain
\beq
\frac{\partial F_k}{\partial y_k}  = y_k \frac{\ud Q}{\ud y} + Q - \frac{\ud C_k}{\ud y_k}.
\label{eqn:partial-deri}
\eeq

Using the optimality condition $\frac{\partial F_k}{\partial y_k} = 0$ in (\ref{eqn:partial-deri}) and solving for $y_k$, we obtain
\beq
y_k = \frac{1}{\frac{\ud Q}{\ud y}} \left(\frac{\ud C_k}{\ud y_k} - Q \right),
\label{eqn:BR}
\eeq
which is in a fixed-point form. Let us denote the function at the right hand side of (\ref{eqn:BR}) by $\mathrm{BR}_k(y_{-k}) \equiv \frac{1}{\frac{\ud Q}{\ud y}} \left(\frac{\ud C_k}{\ud y_k} - Q \right)$, which we referred to as the best response of MNO-$k$ to the action of other competitive sellers. Also, we have $\frac{\ud Q}{\ud y} = -\eta$, and
\[ \frac{\ud C_k}{\ud y_k} = \left\{\begin{array}{ll}
				a_k( p_{\max} + p_c),    & \mathrm{if} \quad 0 \leq y_k \leq (\frac{c_k}{p_{\max}})^{2/\alpha}  \\
				a_k(1 - \frac{\alpha}{2}) c_k y_k^{-\alpha/2} + a_k p_c,  & \mathrm{if} \quad  y_k \geq (\frac{c_k}{p_{\max}})^{2/\alpha}.
			 \end{array} \right.\]
We see that the marginal cost of MNO-$k$ is constant up until a certain point, after which the marginal cost starts to monotonically increase. Thus, the action of MNO-$k$ to sell $y_k$ amount of infrastructure depends on the action of other MNOs, as given by the equation $y_k = \mathrm{BR}_k(y_{-k})$. Substituting $\frac{\ud C_k}{\ud y_k}$, $\frac{\ud Q}{\ud y}$, and $Q$ in (\ref{eqn:BR}),
\beq y_k  = \left\{\begin{array}{ll}
				U_k - y, & \mathrm{if} \quad 0 \leq y_k \leq (\frac{c_k}{p_{\max}})^{2/\alpha} \\
				V_k y_k^{-\alpha/2}+ W_k - y,  & \mathrm{if} \quad y_k \geq (\frac{c_k}{p_{\max}})^{2/\alpha},
			 \end{array} \right.
\label{eqn:BRk}
\eeq
where $U_k = \frac{a_k (p_{\max} + p_c)-\theta}{-\eta}$, $V_k = \frac{a_k (1-\frac{\alpha}{2})c_k}{-\eta}$ and $W_k = \frac{a_k p_c - \theta}{-\eta}$. Recalling that $y = y_k +y_{-k}$, we obtain the best response of MNO-$k$ as
\beq y_k  = \left\{\begin{array}{ll}
				\frac{U_k}{2} - \frac{y_{-k}}{2}, & \mathrm{if} \quad 0 \leq y_k \leq (\frac{c_k}{p_{\max}})^{2/\alpha}  \\
				\frac{V_k y_k^{-\alpha/2}}{2} + \frac{W_k}{2} - \frac{y_{-k}}{2},  & \mathrm{if} \quad y_k \geq (\frac{c_k}{p_{\max}})^{2/\alpha}.
			 \end{array} \right.
\label{eqn:BRk_final}
\eeq
Since $y_k \in [0,1]$, the best response of MNO-$k$ is clipped at (i) $\mathrm{BR}_k(y_{-k}) = 0$ if $\mathrm{BR}_k(y_{-k}) < 0$ or (ii) $\mathrm{BR}_k(y_{-k}) = \lambda_k$ if $\mathrm{BR}_k(y_{-k}) > \lambda_k$.  The status of the market is known from the global price information $Q$ of the infrastructure in (\ref{eqn:revenueQ}). Thus, the actions of the MNOs will be reflected in the market price. An MNO adjusts its action according to the market price as given by its best response function (\ref{eqn:revenueQ}). In (\ref{eqn:revenueQ}), the price $Q$ will be fixed in a given iteration. Thus, it is not necessary for each MNO to know the strategy of other MNOs.

\subsection{Equilibrium of Market}
The equilibrium solution of the Cournot-Nash oligopoly market, $\mathbf{y}^*$, is the fixed point of the best response function. As such, the best responses of all $K$ seller MNOs can be expressed in vector form as $\mathbf{y}^{*} = \mathrm{BR}(\mathbf{y}^{*})$, where $\mathbf{y^{*}} = [y_1^{*}, y_2^{*}, \ldots, y_K^{*}]^{T}$ and $\mathrm{BR}(\tbf{y}^{*}) = [\mathrm{BR}_1(y_{-1}^{*}), \ldots, \mathrm{BR}_K(y_{-K}^{*})]^{T}$. The $[.]^{T}$ denotes transpose of vector.  By taking summation of (\ref{eqn:BRk_final}) over all $K$ seller MNOs, and using the fact that $\sum_{k \in \Kcal \backslash \{0\}} y_{-k}^* = \sum_{k \in \Kcal \backslash \{0\}}(y^* - y_k^*) = (K-1)y^*$, we can  solve the equilibrium quantity $y^{*}$ as
\beq
y^* = \frac{1}{K-1}\sum_{k \in \Kcal \backslash \{0\}} y_{-k}^*.
\label{eqn:Equi_quan}
\eeq
Once the equilibrium quantity $y^*$ is computed, we can find the corresponding equilibrium price $q^*$ by substituting $y^*$ into the price function in (\ref{eqn:revenueQ}), and we obtain $q^* = Q(y^*)= \theta - \eta y^*$.

The equilibrium quantity and equilibrium price of the entire market can be calculated by using \tbf{Algorithm~\ref{alg:Equi_sellers_buyer}}. First of all, we consider the competition among seller MNOs. Once we compute the Cournot-Nash equilibrium for the sellers and obtain $y^*$ and  $q^*$, the buyer MNO-$0$ will use $y^*$ and $q^*$ to compute its best response which is given by \tbf{Algorithm~\ref{alg:GreedyAlgo}}.

\begin{algorithm}
\caption{Market Equilibrium Quantity and Price Between Sellers and Buyer}
\label{alg:Equi_sellers_buyer}
 \begin{algorithmic}[1]
 \STATE Consider competition among seller MNOs.
 \STATE Compute the Cournot-Nash equilibrium of each seller MNO-$k$, $y_1^*,y_2^*, \ldots, y_k^*$.
 \STATE Calculate $y^*$ from (\ref{eqn:Equi_quan}) and $q^*=Q(y^*)$ from (\ref{eqn:revenueQ}).
 \STATE Substitute $y_k^* \to \lambda_k$  and $q^*$ in (\ref{eqn:mincost_case2}).
   \STATE Use \tbf{Algorithm~\ref{alg:GreedyAlgo}} to find the optimal $x_k^*$ and $\Ncal$.
 \STATE Compute $P_c$ from (\ref{eqn:Pc_approx_case2}).
 \end{algorithmic}
 \end{algorithm}

\tbf{Remark:} When multiple buyer MNOs are considered, the buyer MNOs can use the same infrastructure of the seller MNOs without affecting each other's performance because the buyer MNOs use their own spectrum that are orthogonal to each other. In this sense, there is no competition among the buyers, as is assumed in our model. The buyers will only decide how much infrastructure  they need to buy at the given selling price and their required QoS.

\subsection{Stability of Nash Equilibrium of Cournot Oligopoly}
The stability properties of the Cournot oligopoly game are discussed in \cite{Hahn1962, Jesus1980}. Considering Assumption A and Assumptionn B.1 in \cite{Jesus1980}, the sufficient conditions for our system to reach the Nash Equilibrium, which also implies stability of the equilibrium are given as follows:

\begin{proposition}
The Cournot-Nash equilibrium of the oligopoly game is always stable, i.e. the iterative updates of each seller best response always converge to the Cournot-Nash equilibrium, since the following conditions are always satisfied.
\beq
\mathrm{Condition \; 1.} \qquad \frac{\ud Q(y)}{\ud y} - \frac{\ud^2 C_k(y_k)}{\ud^2 y_k} < 0 \quad \forall y_k, Q, k
\label{eqn:con_Nash1}
\eeq 
\beq
\mathrm{Condition \; 2.} \qquad \frac{\ud Q(y)}{\ud y} - \frac{y_k \ud^2 Q(y)}{\ud^2 y} < 0 \quad \forall y_k, Q, k.
\label{eqn:con_Nash2}
\eeq 
\end{proposition}

\begin{IEEEproof}
For Condition 1, from (\ref{eqn:revenueQ}) we can obtain $\frac{\ud Q(y)}{ \ud y} = - \eta$. The second order partial differentiate of $C_k(y_k)$ in (\ref{eqn:Ck}) with respect to $y_k$ is $\frac{\ud^2 C_k(y_k)}{\ud^2 y_k} = a_k \frac{\ud^2 S_k(y_k)}{\ud^2 y_k}$ while using the fact that $y_k = \lambda_k z_k$ in (\ref{eqn:Ck}) and $\frac{\ud^2 S_k(\lambda_k)}{\ud^2 \lambda_k}$ from (\ref{eqn:seconddiff_Sk}). We can express (\ref{eqn:con_Nash1}) as
\beq
\frac{\ud Q(y)}{\ud y} - \frac{\ud^2 C_k(y_k)}{\ud^2 y_k} = \left\{\begin{array}{ll}
                                              -\eta , & 0 \leq \lambda_k \leq   (\frac{c_k}{p_{\max}})^{2/\alpha}  \\
                                              -\eta - a_k \frac{\ud^2 S_k(\lambda_k)}{\ud^2 \lambda_k}, & \lambda_k \geq (\frac{c_k}{p_{\max}})^{2/\alpha}.
			 \end{array} \right.
\eeq
When $0 \leq \lambda_k \leq   (\frac{c_k}{p_{\max}})^{2/\alpha}$, it is always true since $\eta > 0$ hence $-\eta < 0$. Also, when $\lambda_k \geq (\frac{c_k}{p_{\max}})^{2/\alpha}$, where $-\eta <0$, $a_k >0$ and from the proof in (\ref{eqn:seconddiff_Sk}),  $\frac{\ud^2 S_k(\lambda_k)}{\ud^2 \lambda_k} > 0$; hence, the inequality  $-\eta - a_k \frac{\ud^2 S_k(\lambda_k)}{\ud^2 \lambda_k} < 0$ is always satisfied. For Condition 2, since $\frac{\ud^2 Q(y)}{\ud^2 y} = 0$ and $\frac{\ud Q(y)}{\ud y} = -\eta < 0$, the inequality in (\ref{eqn:con_Nash2}) is always true.
\end{IEEEproof}

%

\section{Numerical Results}\label{section:Numerical Results}
We assume that, for all $K+1$ MNOs, the BSs are spatially distributed according to a homogeneous PPP inside a circular area of $500$ meter radius. The seller MNOs are assumed to have the same intensity of BSs per unit area. The maximum transmit power of each BS is $p_{\max} = 10$ dBm, the SINR threshold at each user is $T = 20$ dB, the path-loss exponent is $\alpha = 5$, and noise $\sigma^2 = -150$ dBm. Each BS from all MNOs  transmits at the maximum power in Figs.  \ref{fig:Pc_change_lambda0}-\ref{fig:Asymp_Pc_vs_lambda0}. The price function for selling infrastructure from MNO-$k$, $k \in \Kcal \backslash \{0\}$ in (\ref{eqn:mincost_case2}) is set to $q = \theta - \eta y$, where the fixed cost of installation of infrastructure for all seller MNOs is $\theta = 500$ and the marginal price is $\eta = 5 \pi\times 500^2$.


\subsection{Effect of Changing the Average Number of BSs of MNO-$0$ per Unit Area}

\begin{figure}[h]
\begin{minipage}{0.5\textwidth}
\begin{center}
	\includegraphics[width=\textwidth]{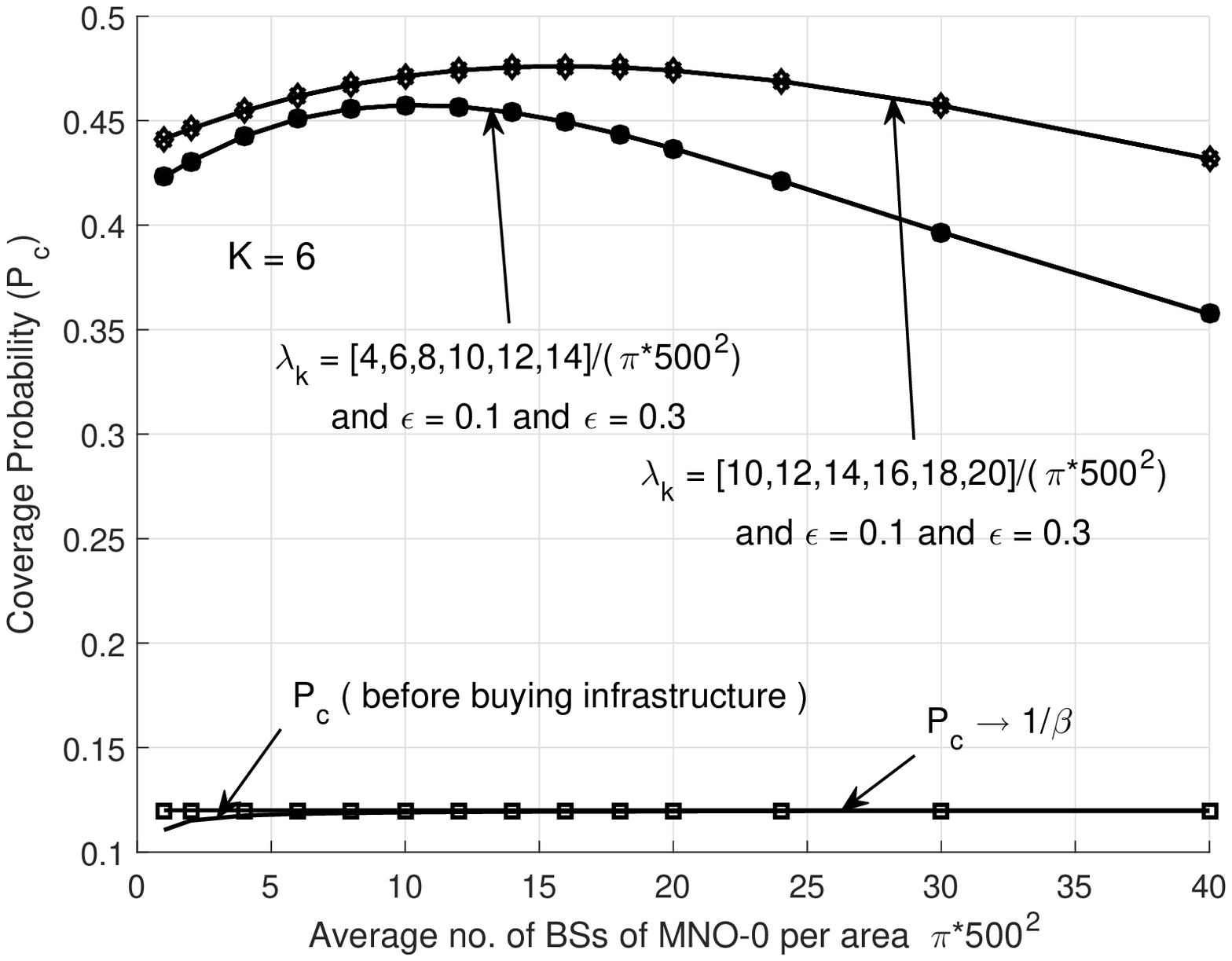}
	\caption{Coverage probability of a user of MNO-$0$ with increasing $\lambda_0$.}
	\label{fig:Pc_change_lambda0}
 \end{center}
 \end{minipage}
\hfill
\begin{minipage}{0.5\textwidth}
\begin{center}
	\includegraphics[width=\textwidth]{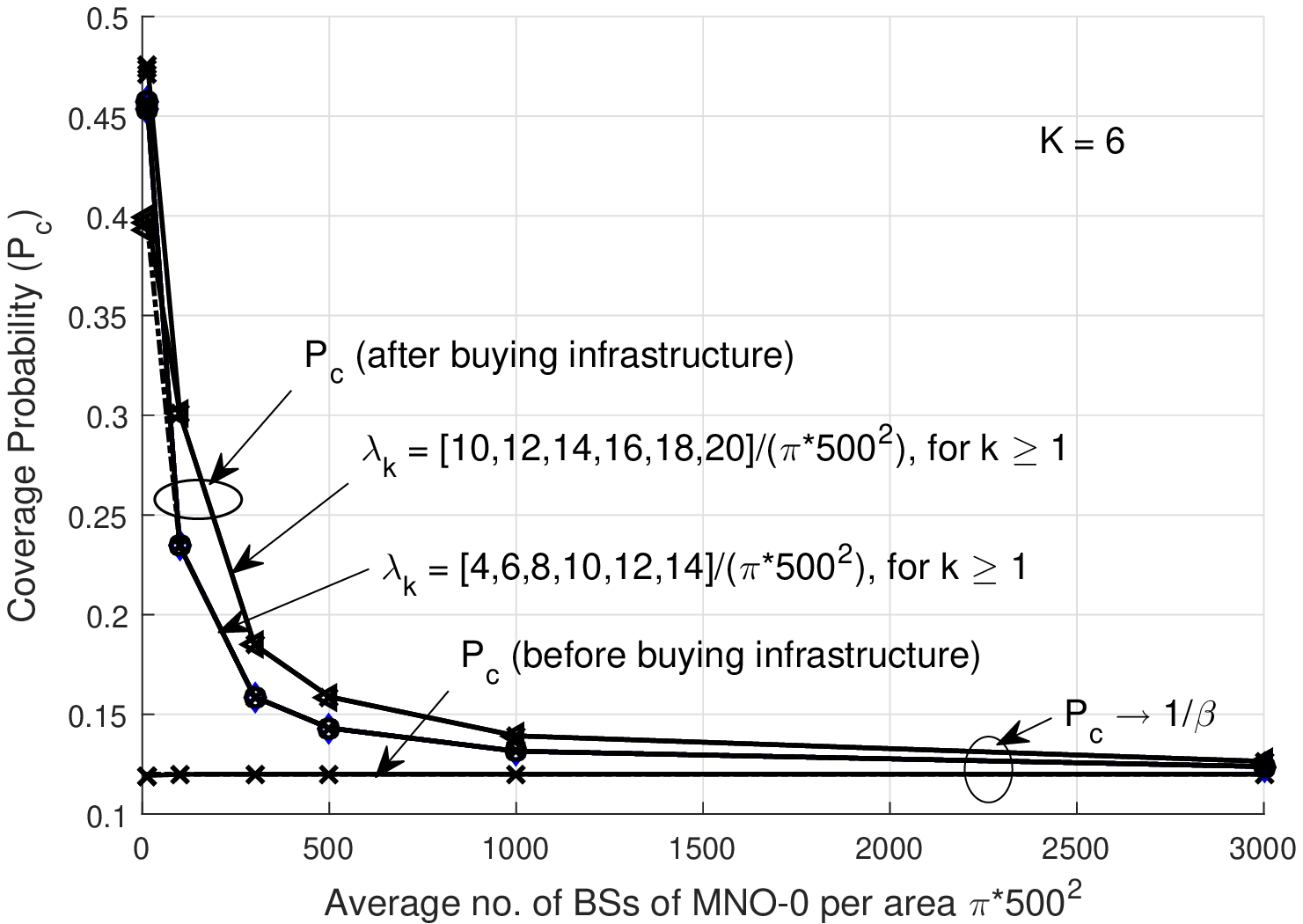}
	\caption{Coverage probability of a user of MNO-$0$ when $\lambda_0 \to \infty$.}
	\label{fig:Asymp_Pc_vs_lambda0}
 \end{center}
  \end{minipage}
\end{figure}


In Fig.~\ref{fig:Pc_change_lambda0}, the difference between coverage probability of a user of the buyer MNO-$0$ before and after buying infrastructure is illustrated. When the MNO-$0$ buys infrastructure, it uses \textbf{Algorithm \ref{alg:GreedyAlgo}} and  the coverage probability is computed using (\ref{eqn:Pc_approx_case2}). When  infrastructure is not purchased, the coverage probability  is computed by using (\ref{eqn:Pc_noinfra}) in \tbf{Proposition ~\ref{prop:Pc_noinfra}}.

When MNO-$0$ uses its own infrastructure, we see that the coverage probability of a user of MNO-0 approaches $1/\beta$ as $\lambda_0$ increases. As such, the MNO-$0$ cannot simply increase its own BS intensity to achieve a coverage more than the upper bound of $1/\beta$. The MNO-$0$ will have to buy more infrastructure to gain more coverage. When the buyer MNO-$0$ buys infrastructure from seller MNOs using \textbf{Algorithm \ref{alg:GreedyAlgo}, the coverage probability of MNO-$0$ is the same for $\epsilon = 0.1$ and $\epsilon = 0.3$. The coverage of MNO-$0$ improves significantly and becomes much greater than $1/\beta$. When the BS intensity of seller MNO-$k$, $k \in \Kcal \backslash \{0\}$ increases, the coverage probability of MNO-$0$ is increased. This verifies \tbf{Proposition \ref{prop:limPc_case2} (ii)}. Also, for fixed $\lambda_k$, where $k \geq 1$, as $\lambda_0$ increases, the coverage of MNO-$0$ decreases, in accordance to \tbf{Proposition \ref{prop:limPc_case2} (i)}. }

In Fig.~\ref{fig:Asymp_Pc_vs_lambda0}, we illustrate the coverage probability a user  of MNO-$0$ before and after buying the infrastructure using a greedy algorithm when $\lambda_0 \rightarrow \infty$. When the MNO-$0$ employs only its own infrastructure, the coverage probability can be computed by using (\ref{eqn:Pc_noinfra}). When the MNO-$0$ buys infrastructure, we assume the number of seller MNOs is six and the tolerable outage probability $\epsilon = 0.1$. It can be seen that $P_c \rightarrow 1/\beta$ when $\lambda_0 \rightarrow \infty$ for both the cases when MNO-$0$ uses its own infrastructure and  when MNO-$0$ buys infrastructure.
When MNO-$0$ buys infrastructure, we only consider the scenario presented by \tbf{Assumption 2.2}, for which the buyer's coverage probability is given by (\ref{eqn:Pc_approx_case2}) and the buyer's strategy given by \textbf{Algorithm \ref{alg:GreedyAlgo}}.  We see that, after buying infrastructure, the coverage probability of a user of MNO-$0$  approaches the bound $1/\beta \sim 0.12$. This verifies the asymptotic analysis in \tbf{Proposition \ref{prop:limPc_case2} (i)}.  

\subsection{Effect of Varying the QoS Parameter}

\begin{figure}[h]
\begin{minipage}{0.5\textwidth}
\begin{center}
	\includegraphics[width=\textwidth]{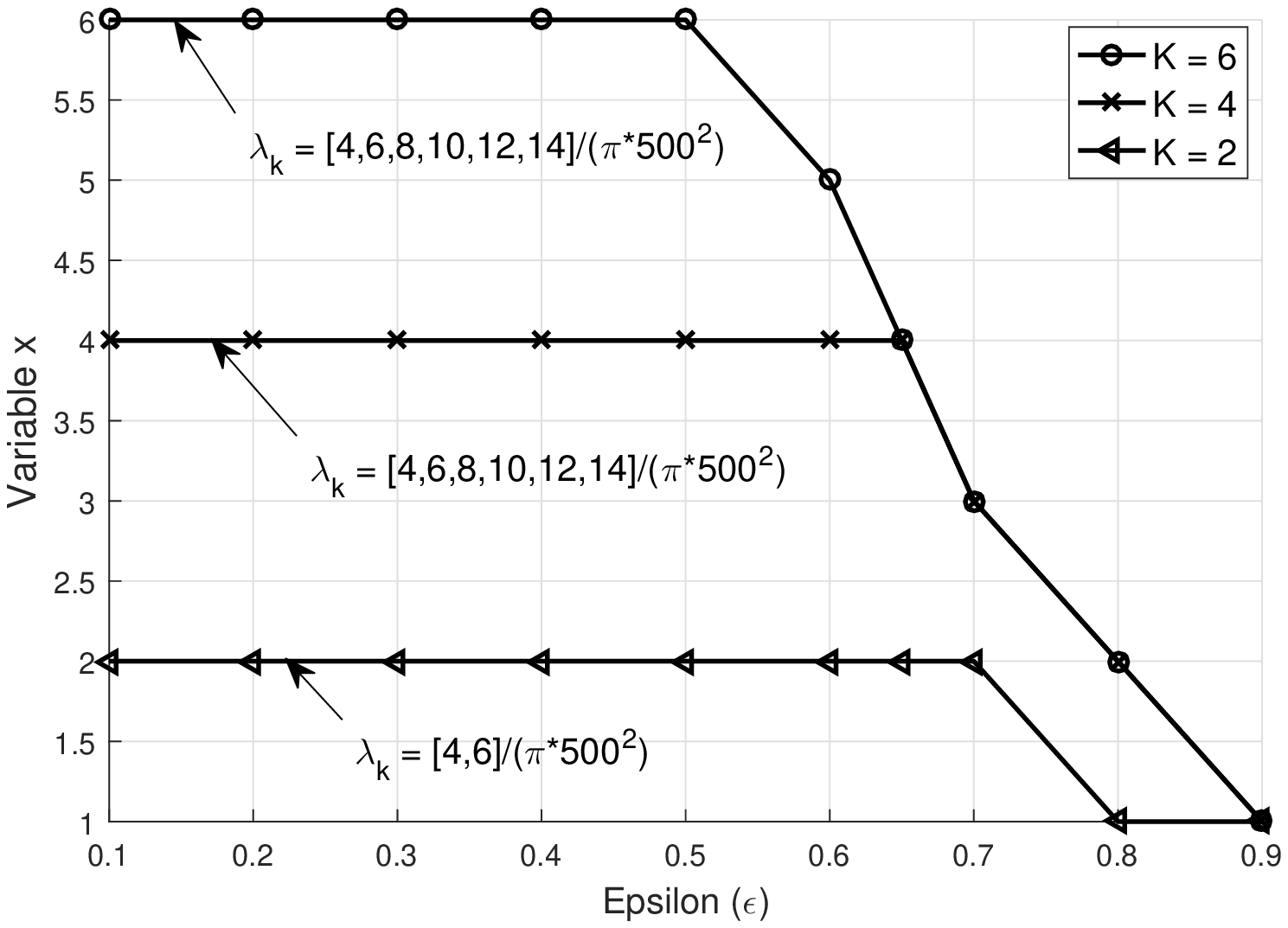}
	\caption{Fractional $x$ for $\sum_{i=1}^N x_i$ versus the tolerable outage probability.}
	\label{fig:fractionx_vs_epsilon}
 \end{center}
 \end{minipage}
\hfill
\begin{minipage}{0.5\textwidth}
\begin{center}
	\includegraphics[width=\textwidth]{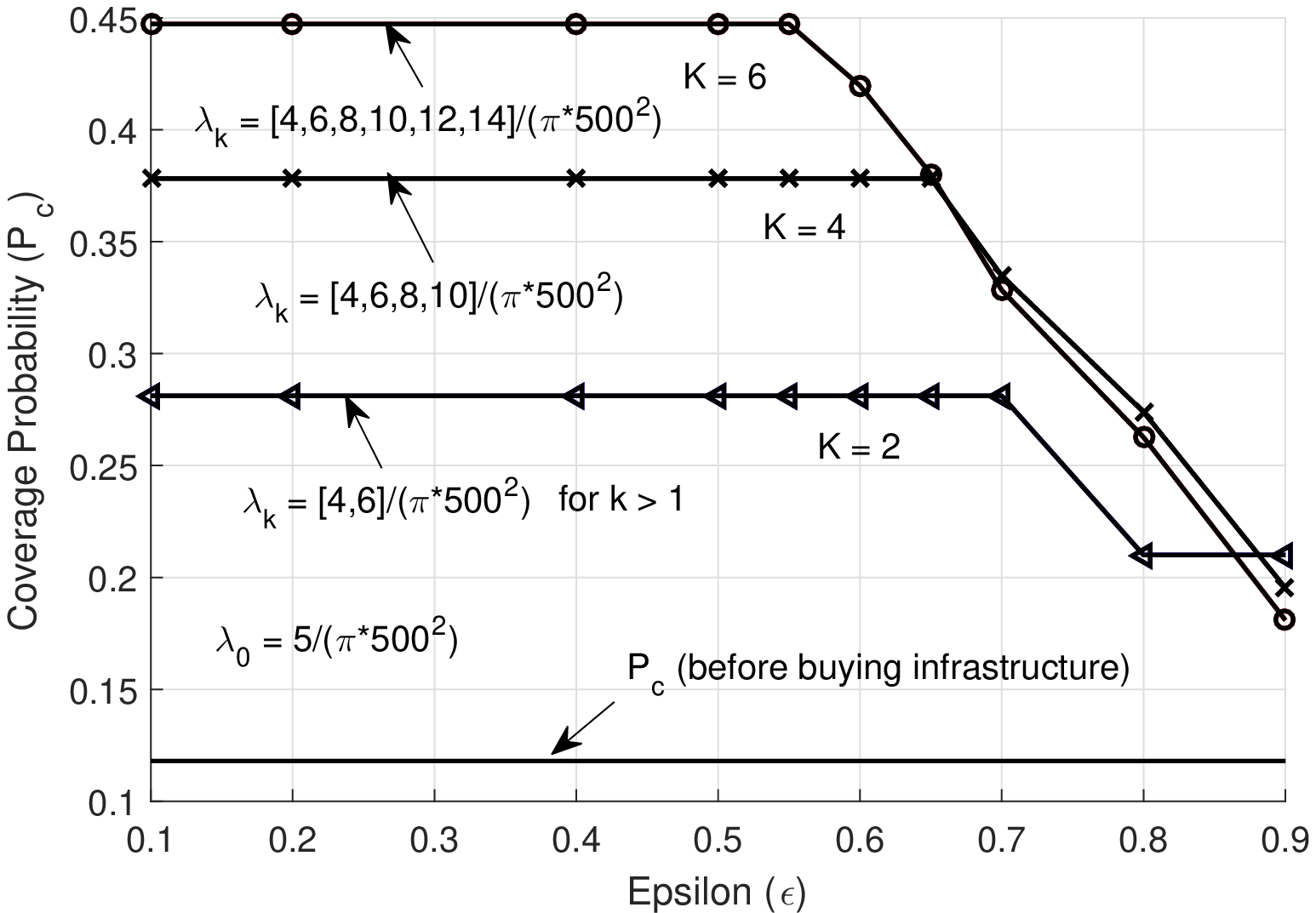}
	\caption{Coverage probability of user of MNO-$0$ when varying the QoS $\epsilon$.}
	\label{fig:Pc_change_epsilon_greedy}
 \end{center}
  \end{minipage}
\end{figure}

In Figs.~\ref{fig:fractionx_vs_epsilon}-\ref{fig:Pc_change_epsilon_greedy}, we show the infrastructure bought by MNO-$0$ while increasing the value of tolerable outage probability $\epsilon$ (i.e. $P_c \geq 1-\epsilon$). We assume that the fixed cost $a$ and the price of infrastructure $b$ for the price function $q_k$ in (\ref{eqn:mincost_case2}) are the same for all seller MNOs. We consider the cases when the buyer MNO-$0$ buys infrastructure from two, four, and six seller MNOs. Fig.~\ref{fig:fractionx_vs_epsilon} plots the fractional values of $x = \sum_{i=1}^K x_i$, which indicates the total proportion of infrastructure that MNO-$0$ purchases as $\epsilon$ changes. We see that as the tolerable outage $\epsilon$ increases, the required amount of infrastructure decreases. For smaller values of $\epsilon$, we see that, in order to satisfy its QoS, the buyer eventually needs to purchase infrastructure from all the available sellers.

The difference between coverage probability of a user of MNO-$0$ before and after buying infrastructure is shown in Fig.~\ref{fig:Pc_change_epsilon_greedy} for varying $\epsilon$. We evaluate the buyer's coverage probability after executing \textbf{Algorithm \ref{alg:GreedyAlgo}} for its purchasing strategy. We observe that for low values of $\epsilon$, MNO-$0$ is unable to satisfy the required QoS solely through its own infrastructure. As shown in Fig.~\ref{fig:fractionx_vs_epsilon}, MNO-$0$ needs to buy from all sellers. However, even after acquiring infrastructure from all the available sellers, it may not be sufficient to satisfy its QoS. As such, for lower $\epsilon$, the coverage probability saturates at a value less than $1-\epsilon$. When $\epsilon$ increases beyond a certain value, $x$ starts to decrease, indicating that at higher $\epsilon$ MNO-$0$ buys less infrastructure.

\subsection{Effect of Varying the Transmit Power from BSs of MNO-$0$ per Unit Area}

\begin{figure}[h]
\centering
\includegraphics[height=3.4 in, width=3.4 in, keepaspectratio = true]{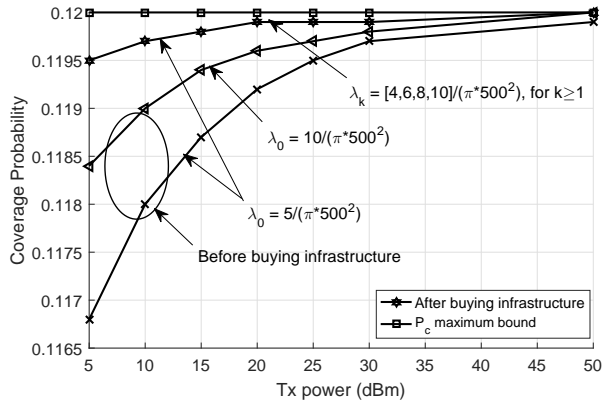}
\caption{Coverage probability of MNO-$0$ when increasing the transmit power from BSs.}
\label{fig:coverage_power}
\end{figure}

Fig.~\ref{fig:coverage_power} illustrates the coverage probability of MNO-$0$ based on the scenario of \tbf{Assumption 2.1}. We plot the coverage probability when the transmit power of BSs of MNO-$0$ increases, while setting the value of $\epsilon = 0.1$. When there is no infrastructure sharing, we consider the cases when $\lambda_0$ is $5/(\pi \times 500^2)$ and $10/(\pi \times 500^2)$. Overall, we see the trend that the coverage improves as the transmit power increases, but quickly saturates to $1/\beta \sim 0.12$. We also see that, for fixed transmit power, when the BS intensity is denser, the coverage of MNO-$0$ is higher.

When the BSs of MNO-$0$ employ low transmit power, MNO-$0$ gains better coverage by sharing infrastructure with the seller MNOs. We plot the case when MNO-$0$ buys infrastructure from four sellers. However, the gain due to infrastructure sharing is very small and does not exceed the upper bound of $1/\beta \sim 0.12$. For all the cases, when the BSs of MNO-$0$ increase their transmit power, the maximum bound of coverage can be achieved easily. There is very little gain for MNO-$0$, in terms of coverage, when buying more infrastructure while its BSs use a high transmit power. This verifies \tbf{Proposition~\ref{prop:power-infra-trade-off-case1}} when $p \to \infty$.

\subsection{Equilibrium Price and Quantity from the Sellers}

\begin{figure}[h]
\begin{minipage}{0.5\textwidth}
\begin{center}
	\includegraphics[width=\textwidth]{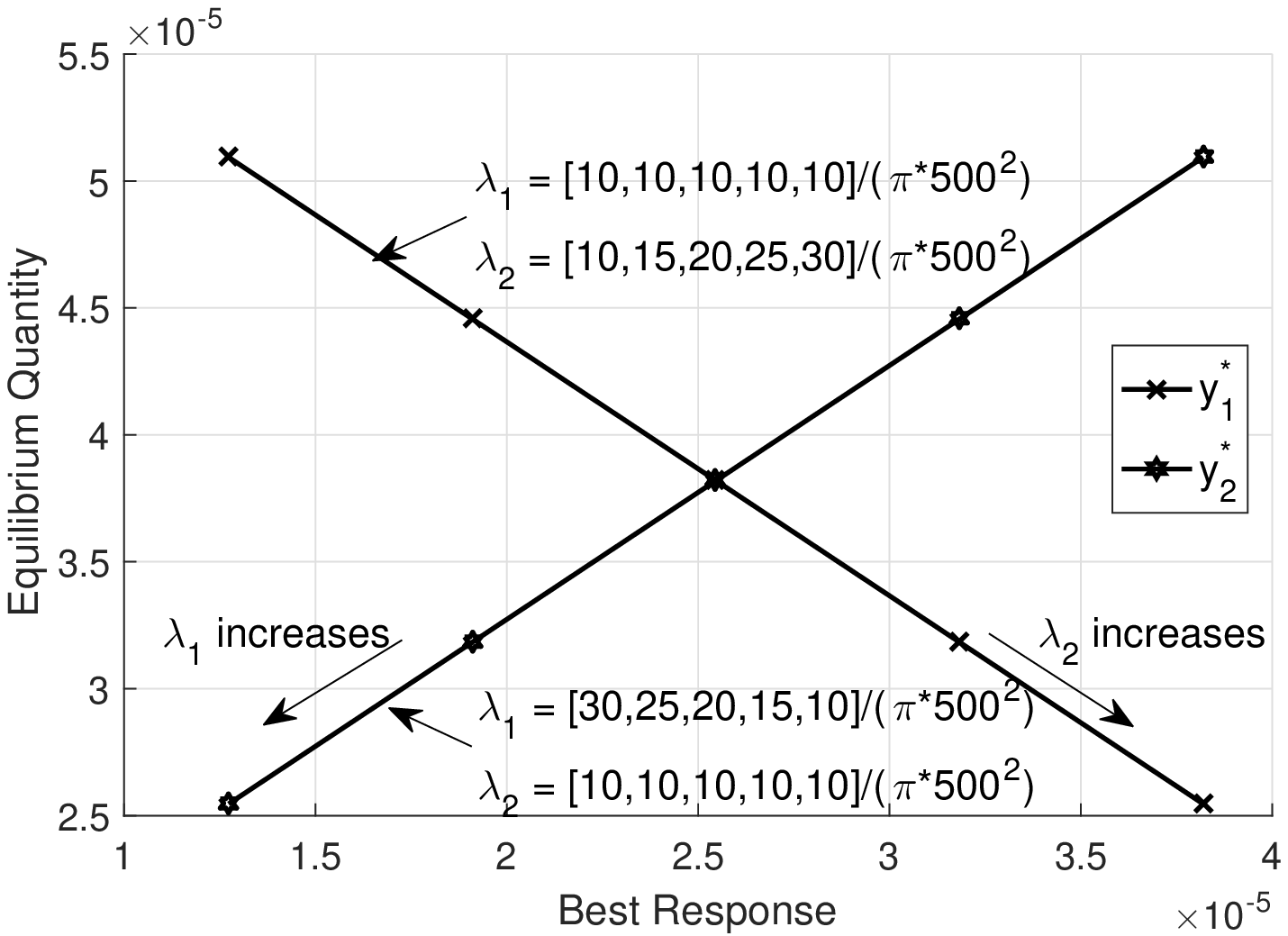}
	\caption{Variation in equilibrium quantity ($y^*$) with $y_1^*$ and $y_2^*$.}
	\label{fig:Equiquantity_y1y2}
 \end{center}
 \end{minipage}
\hfill
\begin{minipage}{0.5\textwidth}
\begin{center}
	\includegraphics[width=\textwidth]{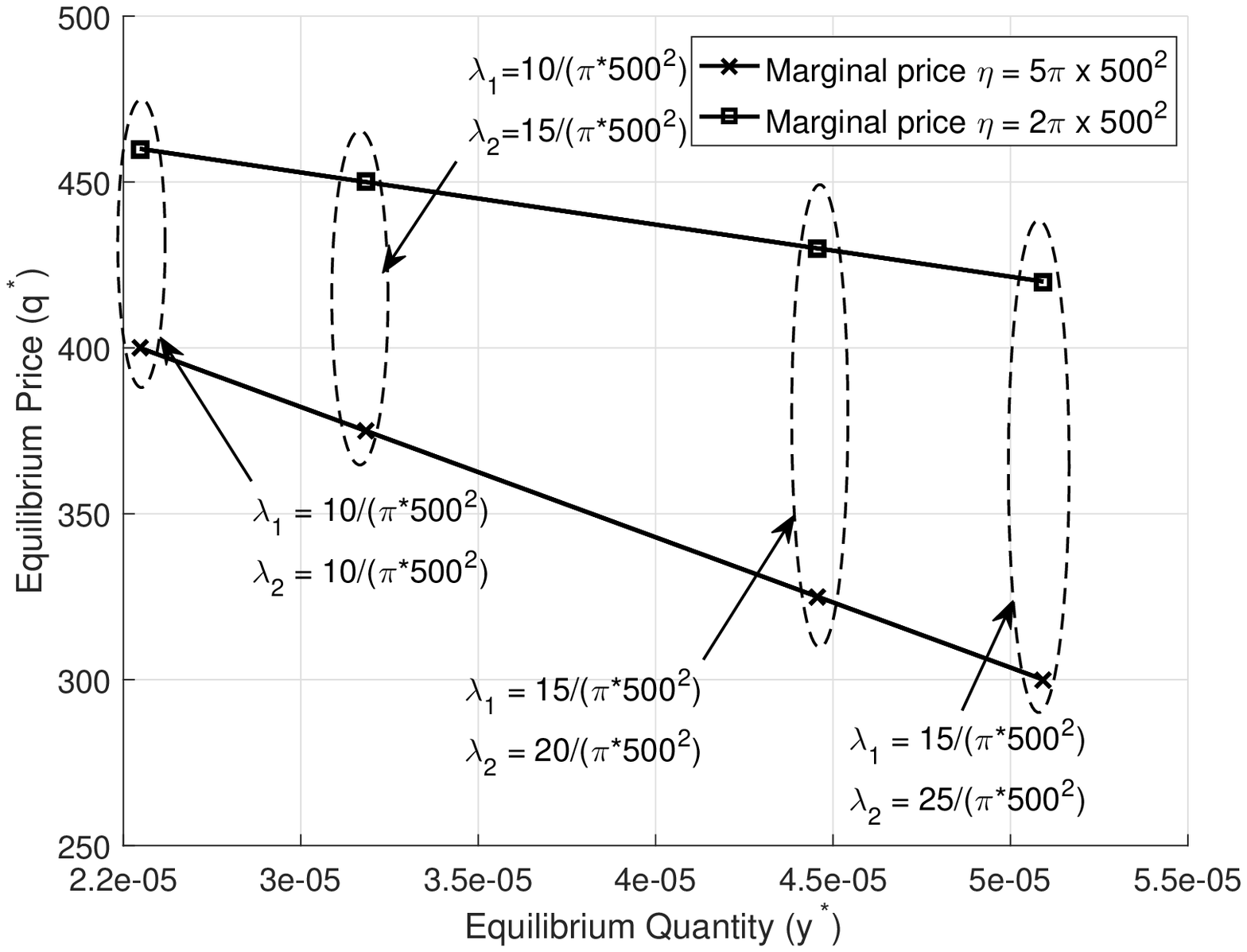}
	\caption{Variation in equilibrium price ($q^*$) with equilibrium quantity ($y^*$).}
	\label{fig:Equipcie_vs_quantity}
 \end{center}
  \end{minipage}
\end{figure}

In Figs.~\ref{fig:Equiquantity_y1y2} and Fig.~\ref{fig:Equipcie_vs_quantity}, we consider the case of two seller MNOs and the simulation parameters are given as follows: the SINR threshold at the user of MNO-$1$ and MNO-$2$ are $T = -15$ dB and $5$ dB, respectively. For the areal power consumption in (\ref{eqn:power-per-area}), we set the fixed circuit power of MNO-$1$ and MNO-$2$ as $p_c = 30$ and $80$, respectively. Also, the cost of power from MNO-$1$ is $a_1 = 50$ and from MNO-$2$ is $a_2 = 90$ in (\ref{eqn:Ck}). As such, it belongs to the case when $y_k \geq (c/p_{\max})^{2/\alpha}$, for $k = {1,2}$ in (\ref{eqn:Equi_quan}).

In Fig.~\ref{fig:Equiquantity_y1y2}, the market equilibrium is illustrated. The best response of seller MNO-$1$ to the action of MNO-$2$ and vice versa can be obtained from (\ref{eqn:BRk_final}). We plot the equilibrium quantity $y^{*}$ from (\ref{eqn:Equi_quan}) with respect to $y_1^*$ and $y_2^*$. Since $y_k = \lambda_k z_k$ where $0 \leq z_k \leq 1$,  the best response $y_k^*$ depends on $\lambda_k$. For fixed value of $\lambda_1 = 10/(\pi \times 500^2)$ and varying $\lambda_2 = [10,15,20,25,30]/(\pi \times 500^2)$, the equilibrium quantity $y^*$ decreases when $\lambda_2$ increases. Similarly, we plot $y^*$ by varying $\lambda_1 = [10,15,20,25,30]/(\pi \times 500^2)$ while setting $\lambda_2 = 10/(\pi \times 500^2)$. The best response of both MNO-$1$ and MNO-$2$ are decreasing functions with respect to $y = y_1 + y_2$. Hence, the best responses of MNOs decrease when $y_1$ and $y_2$ decrease. Also, the best response of MNO-$1$ ($y_1^*$) decreases when $\lambda_2$ increases. Similarly, $y_2^*$ decreases when $\lambda_1$ increases. The intersection point of $y_1^*$ and $y_2^*$ gives a single equilibrium quantity of the Cournot game.

The corresponding equilibrium price $q^*$ is obtained by computing $q^* = \theta - \eta y^*$, which yields a single equilibrium price  for each $y^*$. The equilibrium price  is plotted in Fig.~\ref{fig:Equipcie_vs_quantity} for different marginal prices $\eta$. We consider $(\lambda_1, \lambda_2)  = [(10, 10), (10,15), (15,20), (15,25)]/(\pi \times 500^2)$. When the equilibrium quantity increases, the equilibrium price will decrease due to the law of demand in which the slope of the equilibrium price depends only on the marginal price. While the case of two seller MNOs is considered in Fig.~\ref{fig:Equiquantity_y1y2} and Fig.~\ref{fig:Equipcie_vs_quantity}, the case more than two seller MNOs is given in Fig.~\ref{fig:Pc_Equiprice}.

\subsection{Coverage Probability of MNO-$0$ After Buying Infrastructure at Market Clearing Price}

\begin{figure}[h]
\centering
\includegraphics[height=3.4 in, width=3.4 in, keepaspectratio = true]{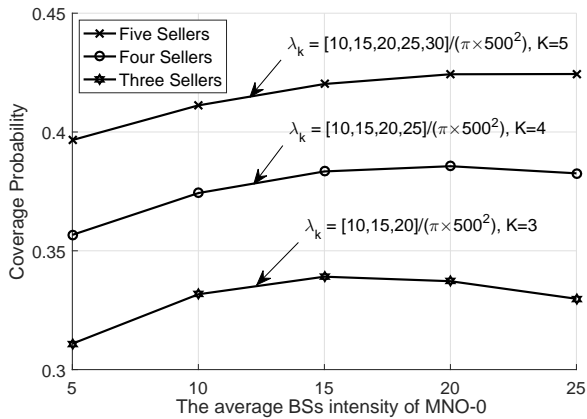}
\caption{Coverage Probability of MNO-$0$  at equilibrium quantity.}
\label{fig:Pc_Equiprice}
\end{figure}

Fig.~\ref{fig:Pc_Equiprice} shows the coverage probability of a user of MNO-$0$ when it buys infrastructure at the equilibrium of market clearing price. By employing  \tbf{Algorithm~\ref{alg:Equi_sellers_buyer}}, all the seller MNOs sell $y^*$ amount of infrastructure at the equilibrium price $q^*$. We consider the cases when the  intensity of BSs of MNO-$0$ is chosen as $[5,10,15,20,25]/(\pi\times 500^2)$, and it buys from three, four, and five seller MNOs. The cost $\theta=500$ and $\eta = 5\pi \times 500^2$. We observe that the coverage of a user of MNO-$0$ improves when there are more  seller MNOs in the market.


\section{Conclusion}\label{section:Conclusion}
We have studied the infrastructure trading problem for multiple seller MNOs and one buyer MNO using stochastic geometry. We have considered two scenarios of infrastructure sharing. Firstly, when \tbf{every} BS of seller MNOs serves a user from the buyer MNO. Secondly, when only \tbf{some} BSs of each seller MNO serves at least one user from the buyer MNO. We have first analyzed the coverage probability of a user of the buyer MNO, and studied the trade-offs between  infrastructure sharing and increasing of transmit power. We have then focused on the strategy selection of the buyer and the competition among sellers. The strategy of a buyer MNO is concerned about how many MNOs and which MNOs to buy infrastructure from in order to satisfy its QoS. The objective of the buyer is to minimize the purchasing cost of the infrastructure. The strategy selection problem of the buyer has been formulated as an optimization problem and the optimal solution was found via Lagrange multiplier method. A greedy algorithm has been proposed to compute the solution. The problem of pricing and finding the fraction of infrastructure that sellers are willing to sell has been formulated using a Cournot-Nash oligopoly game. One of our major conclusions is: infrastructure sharing can improve cellular coverage as long as the interference and association are decoupled.

\bibliographystyle{IEEE}

\end{document}